\documentclass[12pt, twocolumn]{iopart}
\usepackage{epsfig}                                                            
\usepackage{graphicx}
\usepackage{dcolumn}
\usepackage{color}
\usepackage{bm}
\usepackage{subfigure} 
\usepackage{epsfig} 
\begin{document}
\title[Electric field driven destabilization of polaronic insulating state]{\bf Electric field driven destabilization of the insulating state in nominally pure LaMnO$_{3}$}
\author{Rajib Nath$^1$, A. K. Raychaudhuri$^1$, Ya. M. Mukovskii$^2$, Parthasarathi Mondal$^3$, Dipten Bhattacharya$^3$ and P. Mandal$^4$}
\address{$^1$ Department of Condensed matter physics and Materials Science, S.N. Bose National Center for Basic Sciences, Kolkata 700098, India}
 \address{$^2$ National Research Technological University,(MISIS),Leninskii prosp. 4, Moscow, 119049, Russia}
 \address{$^3 $ Nanostructured Materials Division, CSIR-Central Glass and Ceramic Research Institute, Kolkata 700032, India}
  \address{$^4$ Experimental Condensed Matter Physics, Saha Institute of Nuclear Physics, Kolkata 700064, India}
\ead{rajibnath.bu@gmail.com} 
\ead{arup@bose.res.in }
\begin{abstract}
We report an electric field driven destabilization of the insulating state in nominally pure LaMnO$_ {3} $ single crystal with a moderate field which leads to a resistive state transition below 300K. The transition is between the insulating state in LaMnO$_ {3} $ and a high resistance bad metallic state that has a temperature independent resistivity. The transition occurs at a threshold field ($E_ {th} $) which shows a steep enhancement on cooling. While at lower temperatures the transition is sharp and involves large change in resistance but it softens on heating and eventually absent above 280K. When the Mn$^ {4+} $ content is increased by $Sr$ substitution up to x=0.1, the observed transition though observable in certain temperature range, softens considerably. The observation has been explained as bias driven percolation type transition between two coexisting phases, where the majority phase is a charge and orbitally ordered polaronic insulating phase and the minority phase is a bad metallic phase. The mobile fraction $f$ of the bad metallic phase deduced from  the experimental data follows an activated kinetics as $f=f_{o}(E)exp(-\Delta/k_{B}T)$ with the activation energy $\Delta \approx$ 200 meV and the pre-factor $f_{o}(E)$ is a strong function of the field  that leads to a rapid enhancement of $f$ on  application of field leading to the resistive state transition. We suggest likely scenarios for such co-existing phases in nominally pure LaMnO$_ {3} $ that can lead to the bias driven percolation type transition. 
\end{abstract}
\submitto{\JPCM}
\maketitle
\normalsize
\section{\bf INTRODUCTION}
The localization of electronic states due to strong electron-phonon coupling in strongly co-related oxides like $RMnO_ {3} $ ($R$ = rare-earths like La,Pr,Nd etc.) and the effect of external stimulations on these states has been an important research interest over the last two decades \cite{CNR,Tokura}. In material like $RMnO_{3}$, the strong coupling between the electrons and phonons mediated by the Jahn-Teller (JT) distortion, leads to formation of polarons. Below a certain temperature ($T_{JT}$) cooperative JT distortions set in and one obtains a polaronic insulating state accompanied by a orthorhombic distortion of the lattice. Depending on the $R$ ion the value of $T_ {JT} $ can vary between 750K for $R=$ La to 1150K for $R=$ Nd. At a much lower temperature the antiferromagnetic (AFM) order sets in at the Neel temperature $T_{N}\approx 150$K leading to an A-type antiferromagnetic insulating (AFI) state. The physical properties, including crystallographic structure and its temperature evolution through $T_{JT}$  of $RMnO_{3}$ systems like $LaMnO_{3}$ have  been studied in details using various techniques although many important issues still remain unsolved  \cite{Goodenough,Rodriguez,BBvanAken,TChaterji}. The orthorhombic polaronic insulating state with cooperative JT distortion, can be destabilized by hole doping using divalent substitution at the rare-earth site. This leads to insulator-metal transition with long range ferromagnetic order, onset of such phenomena like the colossal magnetorsistance (CMR), charge and orbital ordering (COO) and electronic phase separation, which form the complete gamut of phenomena that make the field of manganites. In addition to the hole doping, the polaronic insulating state can be destabilized by external stimulants like pressure~\cite{Loa}, very high magnetic field \cite{Brion} and as well as by high electric field close to $T_{JT}$ \cite{Dipten}.\\

In this investigation we find that an applied electric field of moderate magnitude can induce instability at room temperature or below in the resistive state of nominally pure single crystal of $LaMnO_{3}$. This leads to a reversible destabilization of the highly resistive insulating state to a lower resistance bad metallic state. (Note: The lower resistance state is called a bad metallic because it has a temperature independent resistivity value which though lower than the insulating state, is higher than those  seen in metallic states of manganites obtained after substantial hole doping). The observed field induced destabilization can be sharp and at lower temperature it can lead to resistance changes of even few orders of magnitude occurring sharply over a small field range. 

The destabilization of the insulating state which we observe in case of nominally pure $LaMnO_{3}$ single crystals  is distinct from the field induced resistance state change (both with memory and without memory) that have been seen in  manganites with much higher level of hole doping (typically $x\geq 0.2$)\cite{CNR2,Tokura2,AyanG1,AyanG}. In these manganite systems with higher level of hole doping (that show Ferromagnetic insulating state or  charge ordered state) the resistive switching has been seen in  single crystals \cite{Himangshu,Aveek} and in films \cite{Rickard Fors} and also in nanoscopic regions that can be created by local probes \cite{Shohini,Jon-Olaf}. Creation of ferromagnetic filaments by high electric field in charge ordered manganites have been reported \cite{AyanG}. Resistive state switching in Schottky junctions involving manganites have also been seen \cite{sawa}. The above list of observation of resistive state switching is not exhaustive and is suggestive. A number of explanations exist for the resistive state transitions in hole-doped manganites. These phenomena range from ionic migration for changes happening near room temperature particularly in junctions \cite{Nian}, electron heating \cite{Himangshu2}, Joule heating \cite{Tokunaga,Jardim}, and electric field induced structural changes \cite{ChJooss,Jon-Olaf}. However, the observed phenomena in nominally pure $LaMnO_{3}$, reported here, is different from  those reported in hole-doped systems. In this case, the starting insulating state is a  polaronic insulating state with Jahn-Teller distorted MnO$_6$ arranged in a cooperative long range ordered state. The observed characteristics in this case have different field and temperature dependences including a sharp field driven transition. The field driven transition appears to change critically for hole doping around $x\approx$0.1. In earlier reports involving hole-doping with concentration $x\geq 0.15$, the starting insulating state was either a charge ordered state (occurring for hole doping $\approx 0.5$) or a ferromagnetic insulating state (occurring for hole doping$\approx 0.15-0.22$). 
\section{\bf EXPERIMENT}
The experiments have been carried out on three nominally pure single crystals of LaMnO$_{3}$ having different dimensions and also on one crystal of Sr substituted La$ _{0.9} $Sr$ _{0.1} $MnO$ _{3} $ to establish the generality of the phenomena and the extent of its sample dependence (see Supplementary). All crystals were grown by Floating zone technique. Sample LaMnO$ _{3}$-1 was grown in Kolkata while samples LaMnO$ _{3}$-2 and LaMnO$ _{3}$-3 were grown in Moscow \cite{Mukovski2}. The La$ _{0.9} $Sr$ _{0.1} $MnO$ _{3} $ crystal was grown in Moscow. To establish the quality of the samples, we have done high temperature powder x-ray diffraction to identify the crystallographic features associated with the cooperative Jahn-Teller transition near $\sim$750K (data given in the supplementary). In addition we have done a series of electron spectroscopies and electron energy loss spectroscopy and Energy Dispersive Analysis of X-Ray (EDAX) to qualitatively ascertain the Mn$^{ 4+}$  content that occurs even in nominally pure LaMnO$_{3}$. We used vacuum evaporated  chrome-gold contact  pads and thin Cu wires for the electrical contacts on the crystals in four-probe configuration (see inset of Figure 2). Resistivity measurement for all the three LaMnO$_{3}$ samples and the La$ _{0.9} $Sr$ _{0.1} $MnO$ _{3} $ sample is done in four probe configuration with low current shown in Figure 1.

 For the measurement of current-voltage characteristics, we used a voltage source and the applied bias ($V_{appl}$) was applied across the sample  in series  with a  standard resistor $R$,  used as a current limiter. The value of series resistor $R$ was selected to be much less than the sample resistance at low bias to ensure that it does not control the current through the circuit. When the sample goes to the low resistive state after the bias driven transition, the current in the circuit increases and it reaches the compliance limit of our measuring instrument. However, with the presence of the current limiter the total resistance of the circuit remains large enough to avoid the compliance limit. We used a Keithley 2410 Source-meter for sourcing the voltage in the circuit and the measuring the current in the circuit. A separate Keithley 2000 DMM was used for the measurement of the voltage drop ($V$) between the two voltage probes of the sample. This arrangement allows us to measure simultaneously the voltage drop across the voltage probes of the sample ($V$) and the applied bias ($V_{appl}$). This arrangement shows the resistive state transition region cleanly. (In the region of resistance measurements where there is no compliance issues involved, the conventional current sourced measurements were used to confirm that the voltage sourced measurements give the same data.) In order to avoid Joule heating, we carried out a pulse mode of measurements with 50$\% $ duty cycle and 400 msec ON period (See Section 3.3). The sample surface temperature was maintained within $\pm$1K during the entire span of recording current-voltage characteristics at each temperature.
\subsection{Resistivity}
 The four probe low current driven resistivity ($\rho$) have been shown in Figure 1.
  \begin{figure}[h]
   \begin{center}
  \includegraphics[scale=0.38]{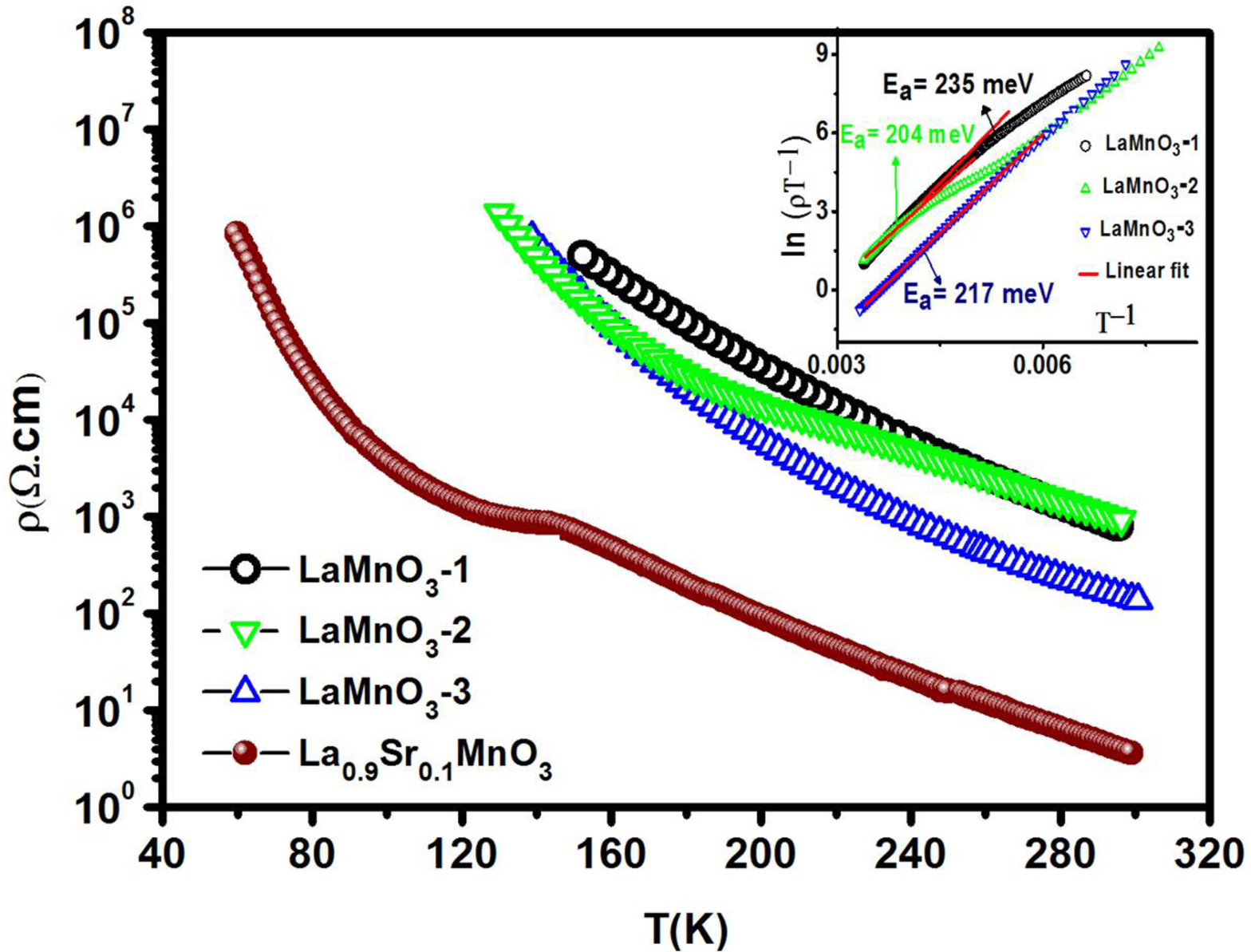}
   \end{center}
  \caption{Resistivities ($\rho$) of LaMnO$_{3} $ and La$ _{0.9} $Sr$ _{0.1} $MnO$ _{3} $ single crystals used in the experiment. The resistivities were measured with very low current in four probe configuartion. Fits of the resistivity data for the LaMnO$_{3} $ crystals to the Adiabaic polaronic model are shown in the inset. The activation energies $E_a$ are noted. For the La$ _{0.9} $Sr$ _{0.1} $MnO$ _{3} $ crystal the ferromagnetic transition occurs at 145K.}
       \end{figure}
  For all the three LaMnO$ _{3} $ samples, the data could be fitted with the adiabatic small polaron hopping model ( see inset of Figure 1 )
   \begin{equation}
  \rho = \rho_{0L}T.exp[-\frac{E_0}{k_B.T}]
  \end{equation}
  Where $\rho_{0L}$ is related to the hopping frequency, length, and dimensionality and E$_0$ is the activation energy. The resistivity data shown in Figure 1 shows the difference in the insulating state of the nominally pure LaMnO$ _{3} $ and the lightly doped LaMnO$ _{3} $, where the presence of finite amount of Mn$^{4+}$ (x=0.1) changes the nature of the polaronic insulating state.  All the three samples of the nominally pure LaMnO$_{3}$ have almost nearly same activation energy  E$_0$ $ \approx $ 217-235 meV (Note: The nonlinearity in the activation energy around 180K for LaMnO$ _{3} $-2 is due to the magnetic inhomogeneity caused by local ferromagnetic moment /cluster formation) and this matches well with previous reported data \cite{Himangshu3}. For La$ _{0.9} $Sr$ _{0.1} $MnO$ _{3} $, E$_{0}\approx$ 145 meV. With the substitution of divalent atom like Ca or Sr in parent LaMnO$ _{3} $,  a hole is created  in the e$_{g}$ levels of the 3d-orbital of Mn leading to  de-localization of carriers  and  increase of the conductivity.
The introduction of Mn$^{4+}$ dilutes the  cooperative JT distortion and enhances the strength of the ferromagnetic Double exchange (DE) interaction. For the Sr doped system, $x$=0.1 region is the critical region\cite{Mukovski}. In this region of hole concentration, for slight variation of Mn$ ^{4+} $ can change the system from highly insulating canted antiferromagnetic (CAF) regime to ferromagnetic insulating (FMI) state at low temperatures. As a result, in La$ _{0.9} $Sr$ _{0.1} $MnO$ _{3} $, there is existence of two distinct phases; Paramagnetic insulating(300-145K) and ferromagnetic insulating (starts below 120K) and in between there is a small metallic region(120K-145K) where $\frac{d\rho}{dT} $ is positive. This would also imply co-existence of complex phases in this region of hole concentration. This complexity makes the x=0.1 region very sensitive to exact Mn$^{4+}$ concentration. Increasing the Sr concentration to $ x > $ 0.15 stabilizes the ferromagnetic metallic state\cite{Renard,Dagotto}. 
        \begin{figure}[h]
        \begin{center}
    \includegraphics[scale=0.39]{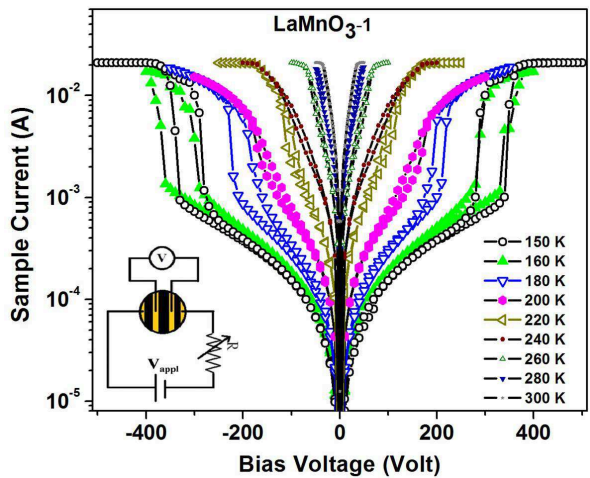}                 
      \end{center}
  \caption {(color on line) Bias voltage (V$ _{appl} $) versus Sample current  characteristics of a LaMnO$_{3}$ single crystal at a few selected temperatures. The inset shows the circuit configuration.}
      \end{figure}
\subsection{I-V data and bias induced destabilization of the polaronic state in LaMnO$ _{3} $}
The $I-V$ data were taken as function of temperature for all the three samples and they are qualitatively similar. Here in Figure 2 we show the data for one sample only and other two data are given in the Supplementary.
The data at low T end were limited by the high resistance of the sample which imposes the compliance limit of the voltage source and the current measuring limit of the Source-meter.
It can be seen that after a particular voltage (which we call the threshold voltage) the current through the the sample switches to a higher value. This arises because the resistive state makes a transition from high resistive state (HRS) to a low resistive state (LRS). The current in the LRS shows saturation due to compliance limit of the source. The LRS prevailed until the voltage was lowered below a threshold value after which it again gets back to the HRS. This transition region shows a hysteresis and it persists up to 260K-280K (depending on the crystal). We have found that both below and above the threshold voltage, the I-V characteristics follow an approximate power law $I\propto V^{n}$. Below the threshold voltage (in the HRS) $n\sim 1$ and above the threshold (in the LRS) $n>$1 (see Supplementary).
      \begin{figure}[!h]
             \begin{center}
              \subfigure [] {\includegraphics[scale=0.35]{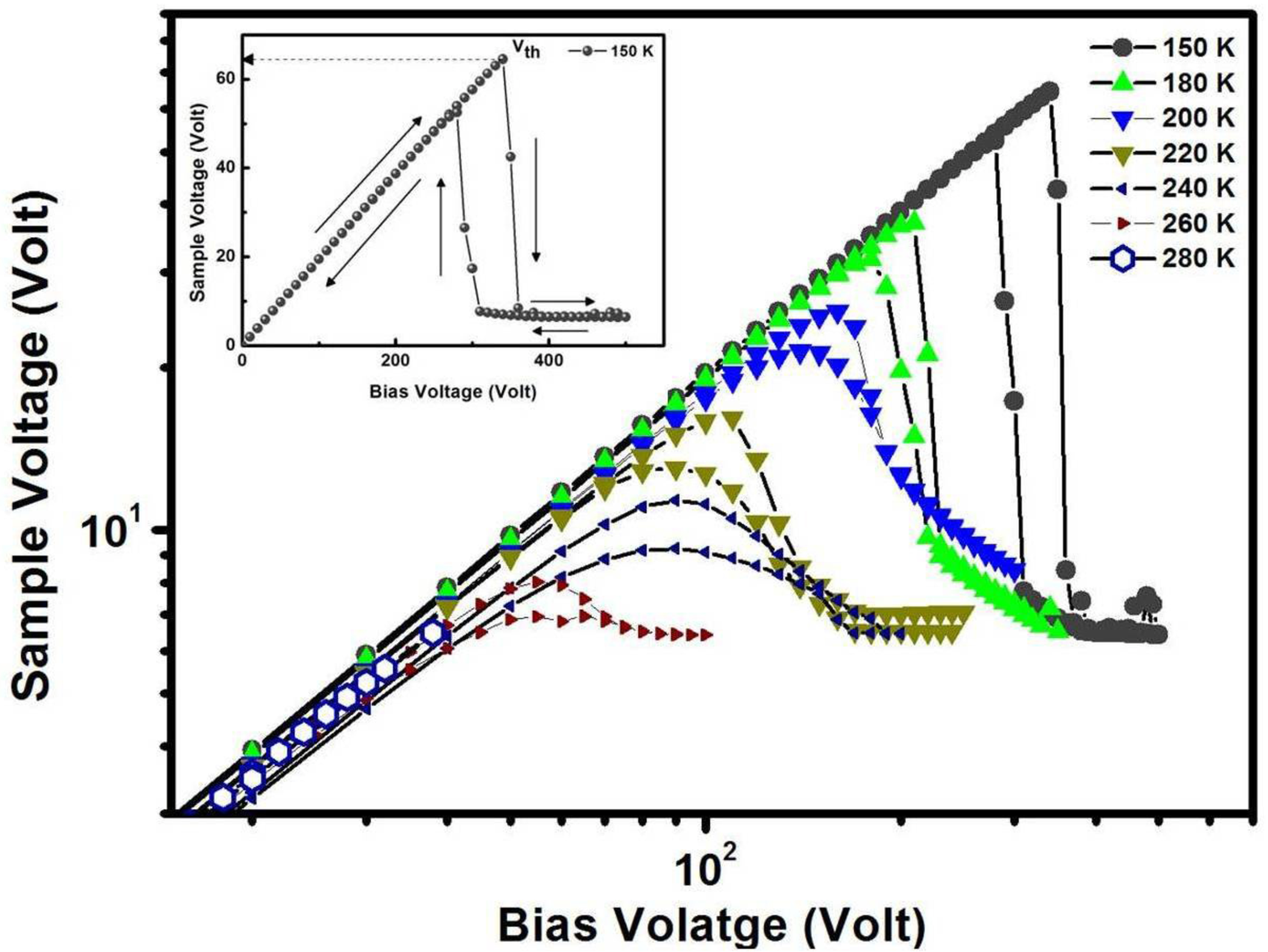}} 
              \subfigure[]{\includegraphics[scale=.35]{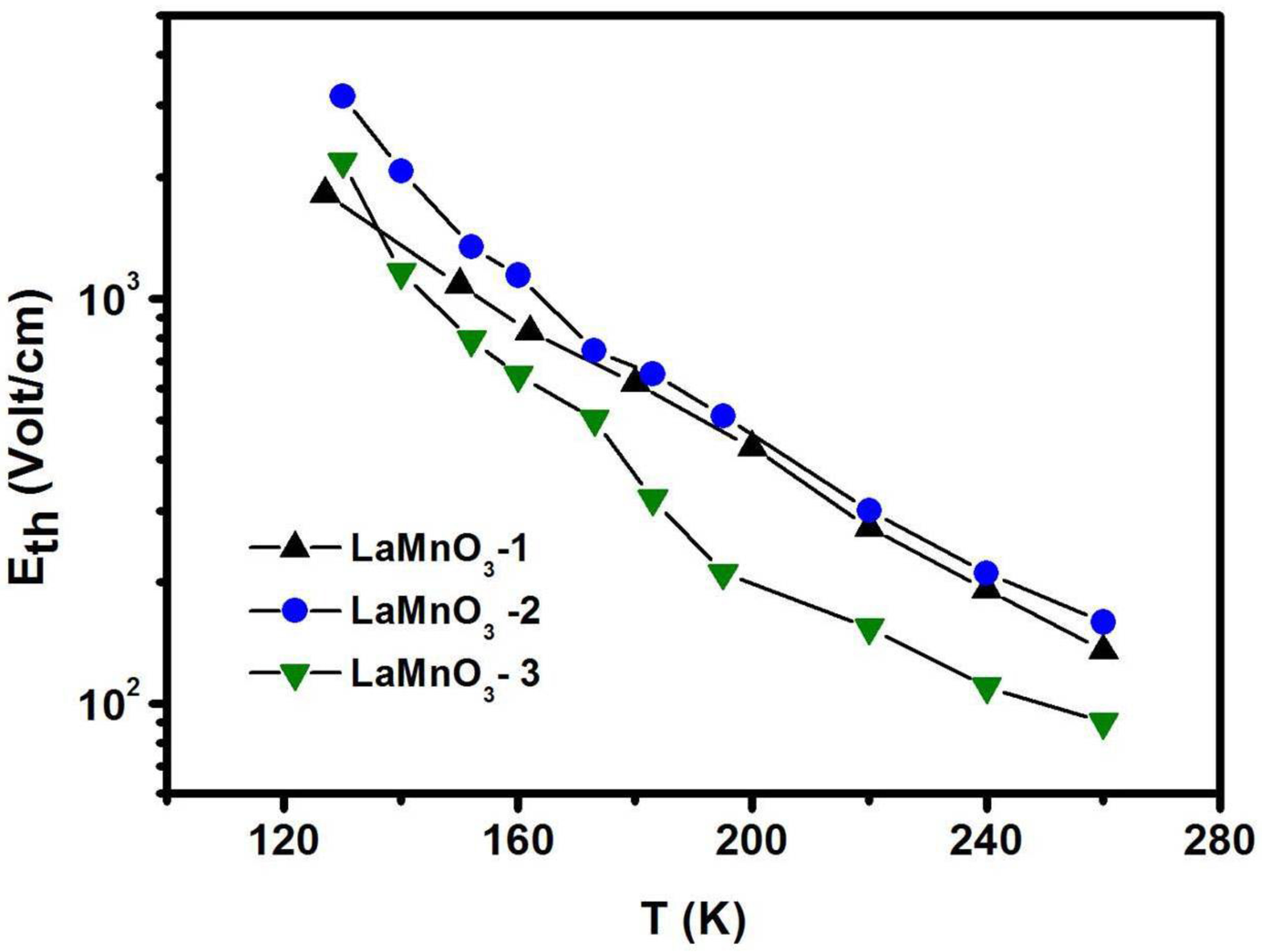}}
             \end{center}
               \caption {(color online) (a) Sample voltage ($V$) versus bias voltage ($V_{appl}$) at different temperatures . Top inset shows a typical threshold voltage $V_{Th}$ at which the sample voltage jumps to a lower voltage value (b)Variation of threshold field E$_{th}$ for the  three different LaMnO$ _{3} $ crystals used.}
             \end{figure} 

In Figure 3(a), we have plotted the sample voltage ($V$) with the applied bias voltage ($V_{appl}$) at different temperatures. This way of plotting the data ($V$ as a function $V_{appl}$), identifies the sample voltage (V$ _{Th} $) at which the transition occurs. When the sample is in HRS (high resistance insulating state) the voltage drop predominantly occurs across the sample and the sample voltage follows the applied bias. On reaching the threshold voltage $V_{th}$, the sample makes transition to the LRS, leading to a fall in the bias across the sample because the bias  now drops predominantly across the limiter $R$. The inset of Figure 3(a) shows an example of the field induced transition and the hysteresis at a representative temperature $T=150K$ and the threshold voltage $V_{th}$ is marked. From this value of $V_{th}$, the  field $E_{th}$ is obtained. In the Figure 3(b) we show the temperature variation of $E_{th}$ for all the three samples. All the samples show a steep rise with reduction in T and have values that are similar (within a factor of 2). $E_{th}$ becomes very small ($ < 100$V/cm.) for temperatures higher than 290-300K, beyond which there is no resistive state transition.
\begin{figure}[h]
  \begin{center}
    \includegraphics[scale=0.39]{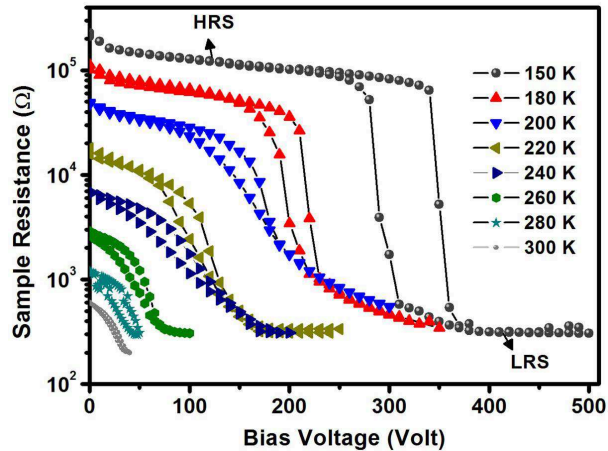} 
    \end{center}
\caption{ Sample resistance versus applied bias voltage; high resistance state is marked as HRS and low resistance state as LRS. The decrease in sharpness of the transition as T increases can be seen. The near temperature independence of the resistance of the LRS can be noted}
\end{figure}

 The change in the resistance at the transition as a function of temperature is shown in Figure 4. The size of the hysteresis region, as well as the jump in the resistance at the transition, reduces with increasing temperature and nearly vanishes after 280-290K. In the HRS the resistivity has strong temperature dependence (see Figure 1). However the LRS has resistance that is nearly temperature independent. \\
 
  The temperature dependences of the HRS and LRS states have been investigated by measuring the resistivities at different constant  bias voltages starting from a very low bias (V$ _{appl} $ $\approx $1V)to a high bias (V$ _{appl} $ $\approx $300V) in four probe method to compare with the resistive state transition as seen in the pulse I-V data in Figure 2. The variation in $\rho$ with T measured at different biases are shown in Figure 5 for one of the LaMnO$ _{3}$ samples (sample-2). For lower biases $<$ 200V (corresponding to E$_{th}$ $<$ E), the sample makes transition to the HRS on cooling below a certain temperature and below 100K we cannot measure the resistivity due to the instrumental measurement limit (marked in the Figure 5). At higher bias ($> 200V$) the resistivity is in the LRS till the lowest temperature 50K due to the steep rise of $E_{th}$. This data (along with the data of Figure 4) clearly demonstrates the existence of a threshold field for the transition and its steep temperature dependence. At higher temperatures when E$_{th}$ is small and applied $E> E_{th}$, the sample is in LRS. On cooling in a fixed $E$, the $E_{th}$ increases sharply and whenever  $E_{th}$ $\geq E$, the transition occurs to the HRS. For the sample shown in Figure 5, there is a clear resitivity jump in the range 125K to 150K. It can be seen that while the resistivity in HRS ($ \rho_{ins} $) has a strong temperature dependence (the polaronic insulating state), the resistivity of the LRS, which is stable for $E_{th}< E$, is nearly temperature independent (marked as $ \rho_{m} $ in Figure 5).\\
  \begin{figure}[h]
   \begin{center}
     \includegraphics[scale=0.40]{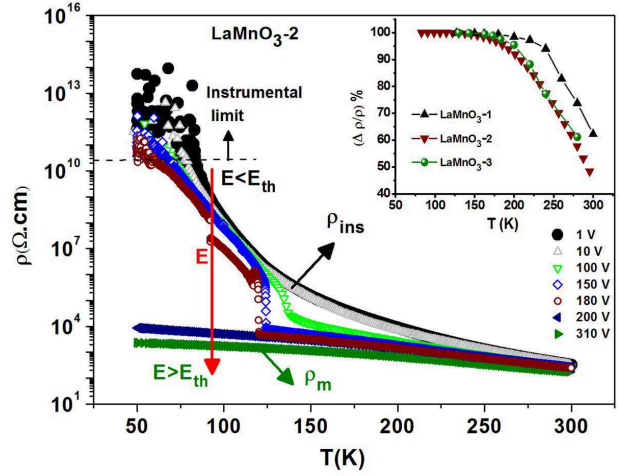}
     \end{center}
 \caption{ Resistivity versus temperature for a LaMnO$_{3}$ single crystal measured  at different constant bias from 1V to 310V in four probe configuration. The transition in resistivity from LRS to HRS on cooling at different bias can be seen.  The inset shows percentage of change in resistivity with temperature for the three samples used at the resistive state transition.}
 \end{figure}
  From Figure 5, it can be seen that $ \rho_{ins} $ equal to or comparable to $ \rho_{m} $ above 280K and there is no observable resistivity transition above 280K. The exact value of this will depend on the exact crystal used. In the inset of Figure 5 we have plotted the percentage of change in resistivity vs. temperature for the three samples. $\frac{\Delta \rho}{\rho}$ is defined as $\frac{\rho_{hrs}-\rho_{lrs}}{\rho_{hrs}}$ (We have used $\rho_{hrs}$ in the denominator and not $\rho_{lrs}$ to avoid unnecessary magnification of the resistance change). It can be seen that the change is nearly 100$\%$ below 210K for all the crystals and it approaches small values for T$ > $280K. While for the other two crystals (LaMnO$ _{3} $-2 and LaMnO$ _{3} $-3 ) the data are identical but there is small quantitative differences near the region where $\frac{\Delta \rho}{\rho}\rightarrow$0 for LaMnO$ _{3} $-1. This probably reflects the differences in the exact $Mn^{4+}$ content of the samples. 
        \begin{figure}[h]
          \begin{center}
      \subfigure[]{\includegraphics[scale=0.35]{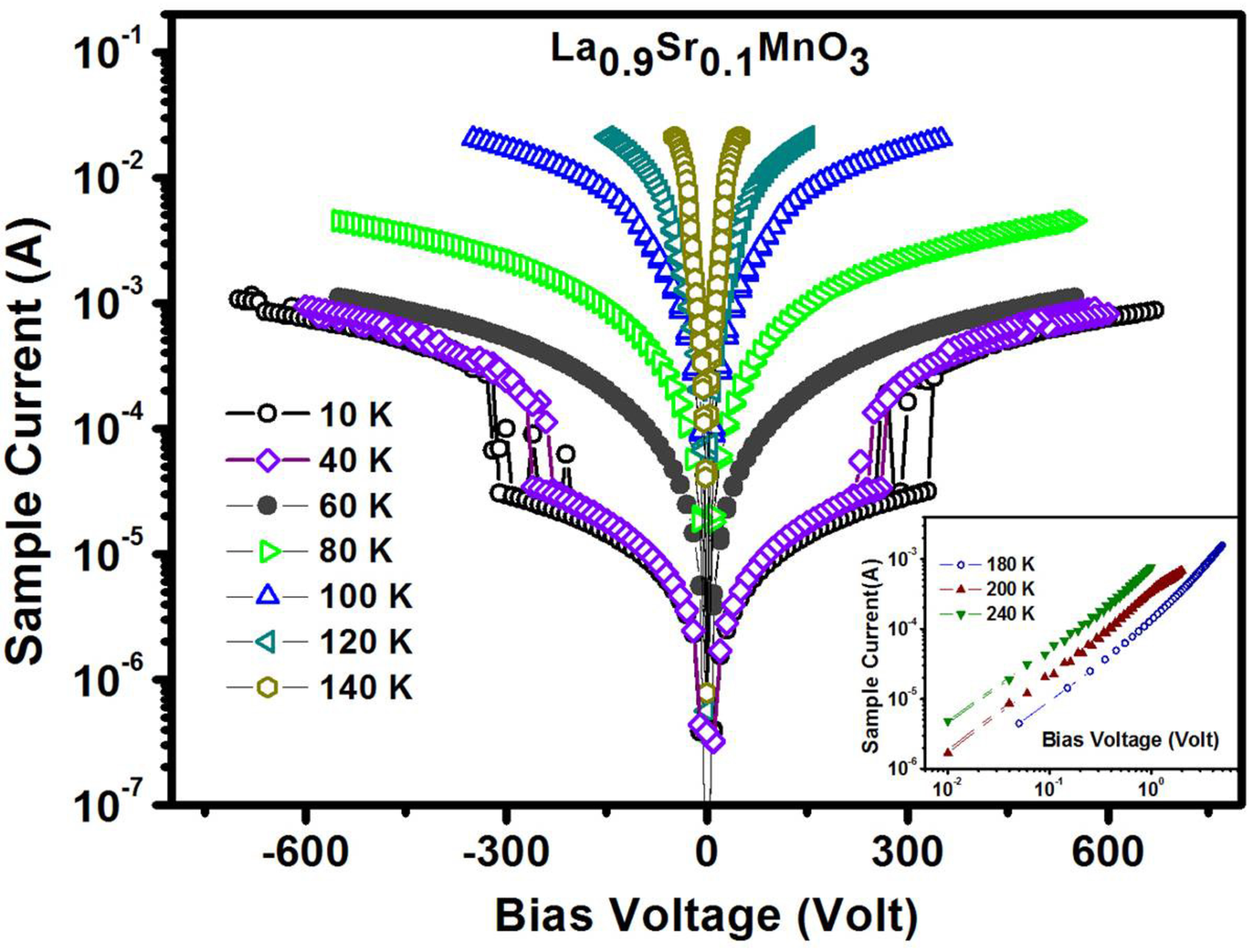}}
      \subfigure[]{\includegraphics[scale=0.35]{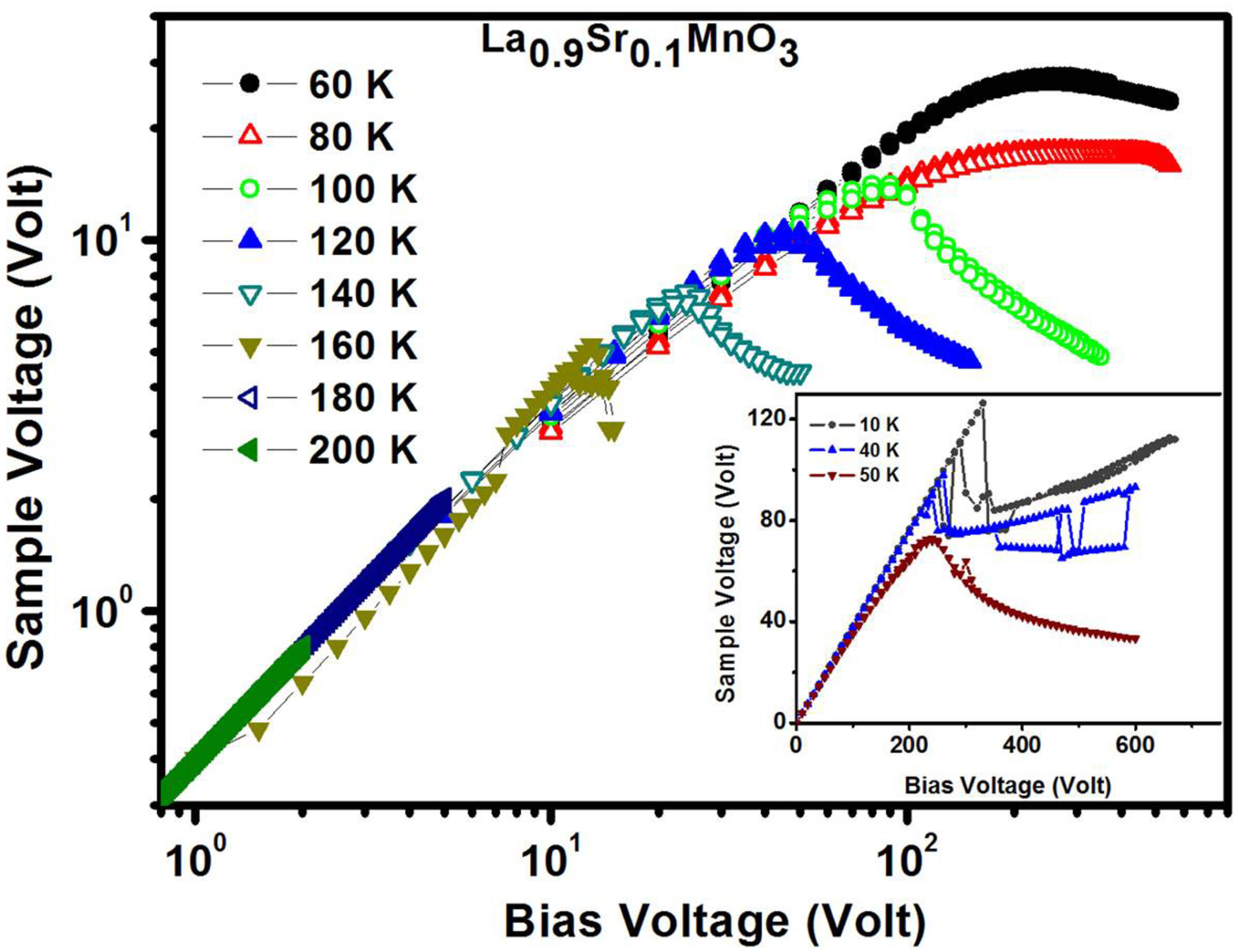} }
            \end{center}
       \caption {(a) I-V characteristics of La$ _{0.9} $Sr$ _{0.1} $MnO$ _{3}$ at different temperatures . Inset shows the data at higher temperatures T $>$ 160K. (b) Sample voltage versus bias voltage at different temperatures in the range 10K $<$ T $<$ 240K. Inset shows the jump of sample voltage at the resistive state transition below 50K.}
        \end{figure}
         
\subsection{I-V data in the hole doped system LSMO (x=0.1) data. }
 To investigate the effect of Mn$ ^{4+} $ on the field driven transition, we have repeated these measurements in La$ _{0.9} $Sr$ _{0.1} $MnO$ _{3}$ (LSMO (x=0.1)) single crystal where the Mn$ ^{4+} $ content is significantly larger than that of a nominally pure LaMnO$_3$. Choice of the doping concentration x=0.1 has been motivated by the fact that it marks the boundary between the CAF state and the FMI state, as stated before. The $I-V$ data in LSMO was taken in the same way as the other crystals (in Figure 6(a)). In LSMO the resistive state transition softens compared to that seen in LaMnO$_{3}$. The resistive state transition occurs at low temperatures below 50K, although the $I-V$ data are strongly non-linear for all T below 140K. The minimum field ($ \sim $ 7 KV/cm) needed to enact the transition is also much larger. 
In Figure 6(b), we plot between $V$ vs. $V_{appl}$ for  10K to 240K. The data at lower temperature $T\leq$ 50K are shown in the inset. For $T\leq $50K, the sample voltage shows similar kind of jump as observed in the three LaMnO$ _{3}$ samples, showing existence of HRS to LRS transition. At higher temperature there is a strong non-linearity leading to a substantial decrease of the sample resistance showing precursor of a resistive state transition that may occur at higher bias. Our data is qualitatively similar as seen by one of us \cite{Mukovski} in LSMO single crystal with nominally same composition. However, there are differences in details that will reflect the difference in exact $Mn^{4+}$ content and also the possibility of co-existence of many phases near the critical concentration x=0.1. The data taken by us as well as by Ref. 26 show that introduction of Mn$ ^{4+}$ gradually changes the nature of the sharp resistive state transition seen in LaMnO$_3$.
\section{\bf DISCUSSION}
The discussion section is divided in two sub-sections. In the first sub-section we describe a simple phenomenological model based on percolation approach to describe the voltage driven resistive state transition. In the next we explore the likely microscopic scenarios for the phenomenological model.
\subsection{A phenomenological model for the transition} 
The model is based on the basic premises that there is a phase coexistence of two phases in LaMnO$_{3}$ below $T_{JT}$, one the orthorhombic polaronic insulating phase with resistivity $\rho_{ins}$ (the majority phase) and a bad "marginally" metallic resistance phase (the minority phase) which has nearly temperature independent resistivity $\rho_{m}$. The rationale for this phase co-existence will be discussed in the next sub-section.
The volume fraction of the minority phase (or the mobile fraction $f$ of it that can contribute to conductivity) at room temperature or below is very small in zero applied bias. As a result the observed $\rho$ is determined by the $\rho_{ins}$. The volume fraction $f$ can be the total volume fraction of the phase with $\rho_{m}$ or better it can represent the "mobile" (or de-pinned) fraction of the phase that will contribute to the  transport. We make this distinction, because some part of the minority phase can be pinned and will not contribute to the transport. This issue will be elaborated in next sub-section. Conducting volume fraction $f$ can be enhanced with applied field $E$ as well as with increase of the temperature $T$. The transition to the LRS occurs when $f$ on application of the field crosses the volume percolation threshold.   

The suggested phenomenological model for the resistive state transition is thus a field induced percolation transition between two phases that coexist below 300K. In this model the applied field does not create any new minority phase, but enhances its mobile fraction $f$. $f$ is also enhanced by the increase of temperature. We evaluated the volume fraction $ f $ from the observed data ($ \rho $ as a function of $E$) in the frame work of effective medium theory \cite{Qingzhong} :  

 \begin{equation}
  \rho_{obs} = \frac{\rho_{ins}\rho_{m}}{(1-f)\rho_{m}+f\rho_{ins}}
  \end{equation}
  \begin{equation}
  f = \frac{\rho_{m}(\rho_{ins}-\rho_{obs})}{\rho_{obs}(\rho_{ins}-\rho_{m})}
  \end{equation}
  
  $\rho_{ins}(T)$ and $\rho_{m}$ are marked in Figure 5.\\
   In Figure 7(a) we show the data of the volume fraction $f$ as a function of $E$ for different temperatures. The $f$ is very small at lower temperatures but it grows very rapidly as the temperature rises. It can be seen that at $E=E_{th}$, $f$ crosses $f_{c}\approx 0.25$, which is close to critical value for volume percolation in 3D, leading to the resistive state transition.
    \begin{figure}[!h]
    \begin{center}
    \subfigure[] {\includegraphics[scale=0.38]{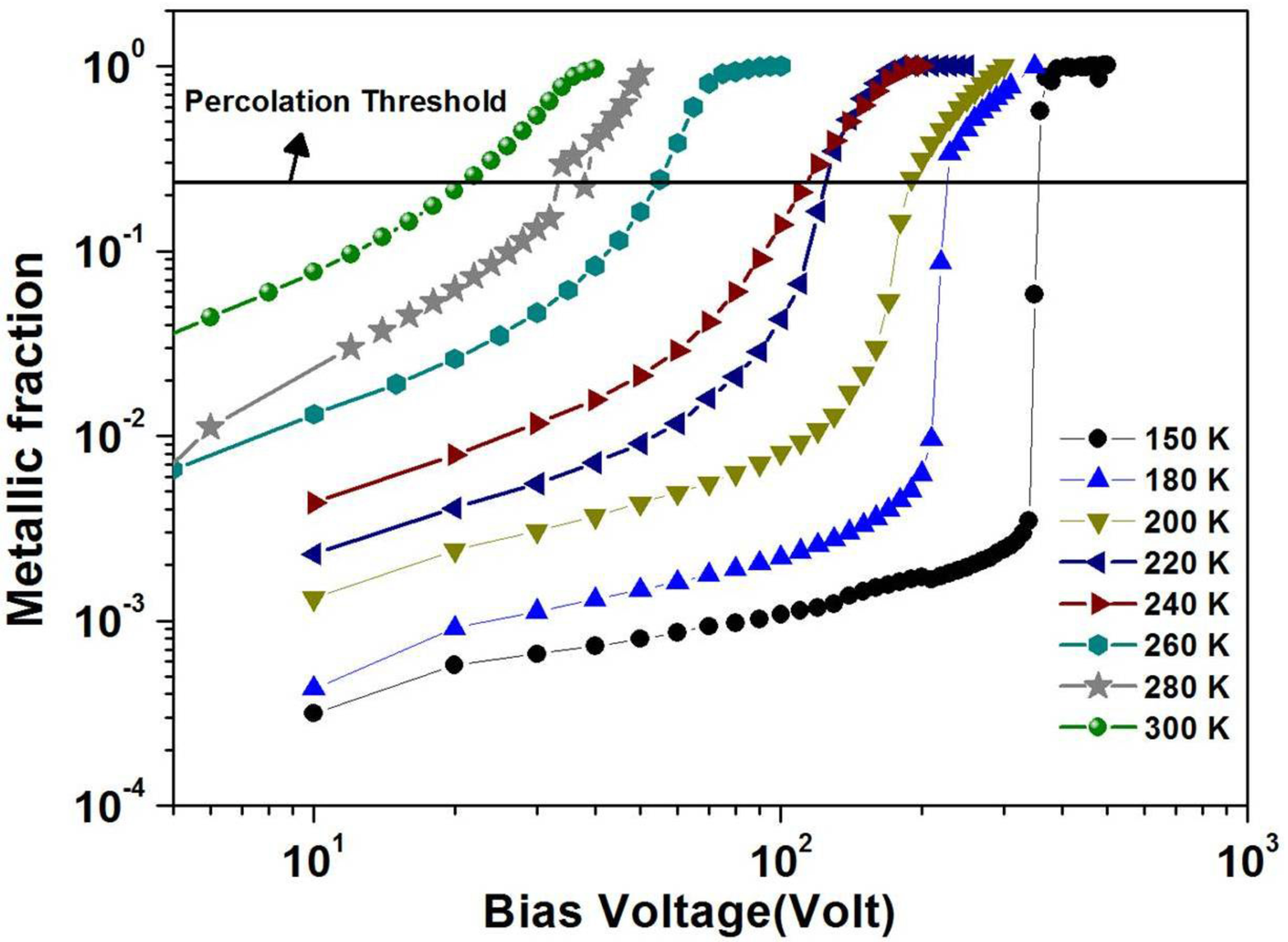}}
    \subfigure[]{\includegraphics[scale=.34]{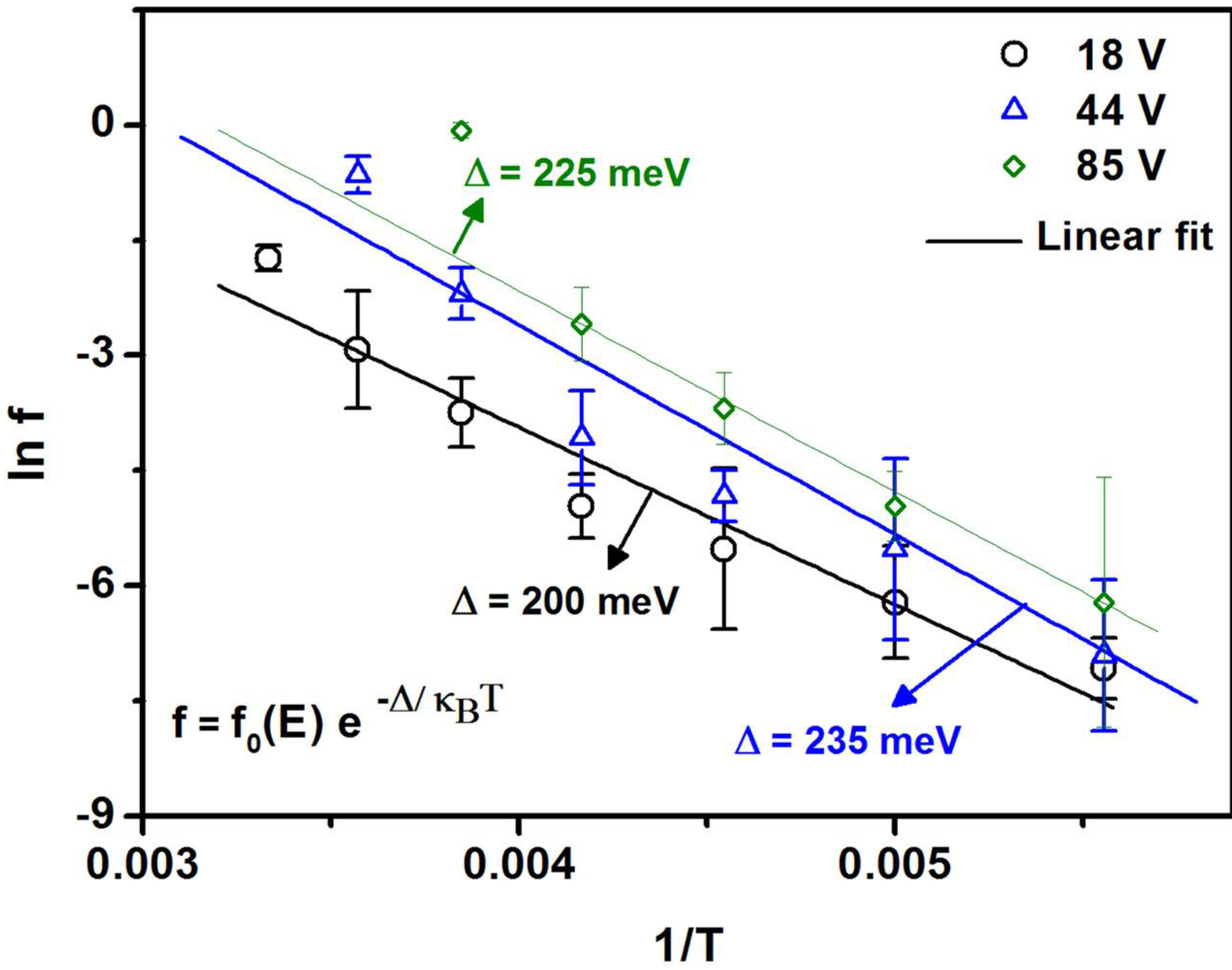}}
    \end{center}
    \caption{(a)Variation of the fraction $f$ with the applied voltage $ V_{appl}$ for LaMnO$_{3}$-1 (b) Activated temperature  variation of the fraction $ f $ . }
    \end{figure}
 From  Figure 7(a) we find that the growth of the volume fraction $f$ is strongly enhanced by an applied field. At lower temperatures for $E=0$, $f$ being very small, the effect of the applied bias is stronger leading to a sharp transition. As the temperature approaches to 280K-290K the decreasing difference between $ \rho _{ins} $ and $ \rho _{m}$ (see Figure 5.) softens the transition considerably.
  
 The transition at a given temperature $T$, in an applied field $E$, occurs when the fraction $f (E,T)$=$f_{c}$. This defines $E_{th}(T)$ at a given temperature $T$. On application of the field, $f$ grows rapidly. As the temperature is raised since $f$ increases even in zero field, the threshold $f_c$ is reached in smaller bias at higher temperature. This ushers in the resistive state transition at smaller bias at higher temperature. The steep dependence of $E_{th}(T)$ on T is thus a reflection of the strong dependence of $f$ on $T$ and $E$, which we show below depends on thermal activation and  field induced de-pinning respectively. 
 
 The temperature dependence of $f$ was found to be an activated process following the relation: $f = f_{0}(E)exp (\frac{-\Delta}{\kappa_{B}T})$, $\Delta$ being the activation energy. This is shown in Figure 7(b) for three representative biases. Thermal activation energy $\Delta$ is found to be in the region of 200-235 meV and largely independent of $E$. Interestingly the value of $\Delta$ is the same as the activation energy ($E_a$) found in the polaronic state in the three LaMnO$ _{3} $ crystals. Interestingly, The pre-factor $ f_{0}(E)$ increases with $E$. We find that the predominant effect of the electric field $E$ is to change the pre-factor $f_{0}(E)$. The dependence of a $f_{0}$ on $E$ can be described by $ f_{0} = C exp(\frac{-E_{0}}{E}) $, where $ C $ is a constant and $E_{0}$ is a scale for the de-pinning field for the pinned minority phase. This is very similar to the relation found in many field driven de-pinning phenomena including charge density wave de-pinning \cite {Bardeen,Farges}. Thus the joint effect of $E$ and $T$ on the volume fraction $f$ is given as:
 \begin{equation}
  f = C exp (\frac{-E_{0}}{E}) exp (\frac{-\Delta}{\kappa_{B}T})
   \end{equation}
To summarize, the observed resistive transition has been described as a bias driven percolation type transition between two states with coexisting phases: the polaronic insulating phase and a lower resistive bad metallic phase. The transition is controlled by two parameters. The first is the mobile fraction $f$ of the more conducting phase that has a steep $E$ and $T$ dependence that is described by eqn. 2. The second parameter is the relative differences between the two phases $\rho_{ins}$ and $\rho_{m}$ as shown in Figure 5. This difference determines the magnitude of the resistive jump at the transition. At $T\approx$ 280K-290K since $\rho_{ins} \rightarrow\rho_{m}$, the transition is not observable above this temperature . 
 
 The creation of $Mn^{4+}$ by Sr substitution in LaMnO$_{3}$ changes the nature of transition and can gradually suppress it. The data on the single crystal La$ _{0.9} $Sr$ _{0.1} $MnO$ _{3}$ taken by us along with similar data in Ref. 26 show that in most of the temperature range the  transport shows strong non-linearity but $ f $ may not exceed $ f_{c} $. The strong non-linearity can be also modelled by the phenomenological model described for the nominally pure LaMnO$_{3}$, where the mobile fraction $f$  gets enhanced by the applied bias leading to the strong non-linearity. In this case the bias dependence of $f$ for most of the temperature range is much less steep compared to the nominally pure system. 
 
From the experimental data using the 2-phase analysis we find the resistivities of the two phases $\rho_{ins}$ and  $\rho_{m}$ for La$ _{0.9} $Sr$ _{0.1} $MnO$ _{3}$.
 \begin{figure}[h]
           \begin{center}
             \includegraphics[scale=0.40]{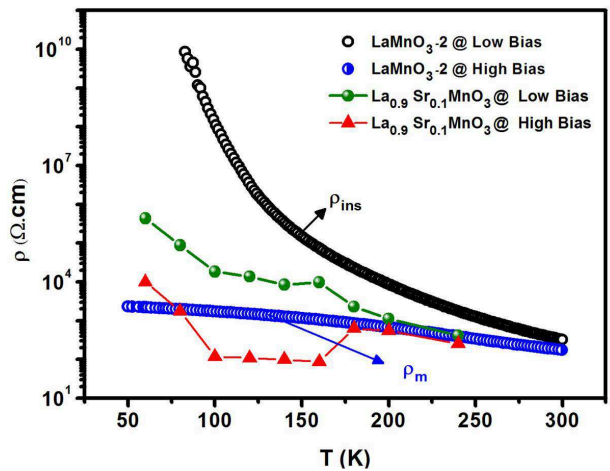} 
             \end{center}
         \caption{Resistivities of LaMnO$_3$ and LSMO (x=0.1) in HRS in comparison to the $\rho_m$ of the LRS as deduced from the experiment.}
         \end{figure}
 In Figure 8 we show the data. The resistivity $\rho_{ins}$ in the  polaronic insulating state is much suppressed by Sr substitution. However, for $T \geq$150K, the resistivity of the minority phase $\rho_{m}$ is identical to that of the value of $\rho_{m}$ that we found for LaMnO$_3$. This is an important observation that while the creation of Mn$^{4+}$ suppresses the resistivity of the polaronic insulating state considerably, the resistivity of the minority bad metallic phase is not affected. This suggests that whatsoever be the origin of the bad metallic phase that give $\rho_{m}$, it is not created by the deliberate hole-doping with $Mn^{4+}$ that is created by Sr substitution. It can be seen that in the temperature range $T \geq$150K, the value of $\rho_{ins}$ for the Sr substituted system being lower it is not much different from the $\rho_{m}$. As a result the resistive transition becomes soft and looses the sharpness seen in the LaMnO$_3$. It shows up as a non-linear $I-V$ curve as $f$ changes with bias.  Below 150K, when the FM state sets in, there is phase co-existence with lower resistance ferromagnetic metallic phase. This brings down $\rho_{m}$, which however, recovers below 100K when the ferromagentic insulating state sets in. In this region $\rho_{ins}$ also rises rapidly on cooling. This leads to a sharp transition at lower temperature. The quantitative details of the transition where one has co-existing phases of different kind would depend on the exact value of the Mn$^{4+}$. 

\subsection{Microscopic scenarios for the phenomenological model}

The phenomenological model given above is based on phase co-existence of a relatively low resistance marginally (or bad) metallic phase that has nearly temperature independent resistivity along with the polaronic insulating phase. This was seen to occur in nominally pure LaMnO$_{3}$ as well as in La$ _{0.9} $Sr$ _{0.1} $MnO$ _{3}$. In this subsection we explore whether there are evidences of such a phase coexistance and a microscopic scenario for such a bad metallic phase.

In  marginally pure LaMnO$ _{3}$ with very small intentional Mn$^{4+}$, the dominant phase below  $T_{JT}$  is an orthorhombic polaronic insulating phase which leads to A-type antiferromagnetic ordering below 150K. However, there are evidences based on Neutron scattering studies \cite{Huang} Resonant X-Ray scattering \cite{Prado} and also from  optical conductivity experiments \cite{Kovaleva} that this picture may not be complete and there are evidences of presence of small concentration of a phase that have similarity in structure and electronic properties with the high temperature phase that is found above $T_{JT}$. This minority phase can be another orthorhombic and rhombohedral ferromagnetic phases with a considerably smaller unit-cell volume with a ferromagnetic transition temperature near $T_N$. Based on the above experiments, it has been recently suggested \cite{Moskvin} that such a minority phase can be a bad metallic phase. Strong evidence of this comes from pressure induced metallization of LaMnO$_3$. It had been observed \cite{Loa} that the pressure induced destabilization of the insulating state in LaMnO$_{3}$ at 300K  (at pressure $\geq $ 32 GPa) leads to an orbitally dis-ordered phase that is like a bad metal with temperature independent resistivity. Subsequent X-ray Absorption Spectroscopy in LaMnO$_3$ under pressure \cite{Ramos} showed that although the insulator-metal is completed above 32 GPa, there is co-existing orbitally dis-ordered phase that exists in a wide pressure range above 7GPa till the transition is complete.
 Thus there is evidence that a bad metallic phase with small volume fraction can co-exist with the insulating phase well below $T_{JT}$ in LaMnO$_3$.

 It has been suggested~\cite{Moskvin} recently  that  charge transfer (CT) instabilities can lead to dynamic charge disproportionation ($Mn^{3+}+ Mn^{3+} \rightarrow Mn^{4+}+Mn^{2+}$) in nominally pure LaMnO$ _{3}$,  which can act as an electron-hole (EH) pair. The presence of this instability leads to phase separation between a Jahn-Teller distorted polaronic insulator (the conventional phase) and a phase that behaves as a bad (or marginal) metal comprising of EH droplets ($Mn^{4+}-Mn^{2+}$ pairs). There are evidences \cite{Murakami,Zimmermann} that such a phase is present well below $T_{JT}$ and even at 200K with a finite volume fraction.  Increase in temperature leads to enhanced volume fraction of this phase and the phase transition at $T_{JT}$ is envisaged to arise from this phase. 

 It is suggestive that the bad metallic minority phase, proposed in the phenomenological model above, may be related to the this CT instability driven EH droplet phase ~\cite{Moskvin}. The full volume of EH droplets that make the minority phase need not be mobile and it can be self-trapped at lattice sites or may also be trapped by local impurity potentials. The volume fraction ($f$) of the minority metallic phase  estimated from the electrical resistance data gives the volume fraction of the mobile phase. The enhancement of $f$ occurs with temperature following an activated process that is very similar to the activation energy seen in transport in the polaronic insulating phase. The electric field applied can lead to de-pinning of the EH droplets from the lattice sites or impurity pinning sites. The nature of the $E$ dependence of $f$ seen by us suggests that this can indeed be the case.
 
 Creation of Mn$^{4+}$ by Sr substitution is likely to be  different from the Mn$^{4+}$ in the EH pair. In a very qualitative sense, Mn$^{4+}$ in EH pair is like intrinsic doping while that created by substitution is like extrinsic doping. With creation of more substitutionally doped Mn$^{4+}$, they will dominate over the intrinsic EH pair seen in nominally pure LaMnO$_3$. The inhibition  of the resistive state transition in La$ _{0.9} $Sr$ _{0.1} $MnO$ _{3}$ can be a manifestation of this. 

\subsection{Issue of Joule heating}
Joule heating have beeen suggested as the sole origin of resistive state switching observed in manganites \cite{Tokunaga, Jardim}. Below we discuss that such is not the case for the observed switching in the parent mangnite samples. To check the contribution of  Joule heating in our ovserved resistive switching, we have done a current level measurement as a function of time for one of the LaMnO$ _{3} $ samples at low temperature with two fixed voltages (in Figure 9). We found nearly a constant sample current level with the time when a bias of 200 V is stepped to 305 V. These are voltages just before and after the resistive state switching. If there is any subtantial Joule heating that has occured in the sample, sample resistance and hence the current in the sample would drift with time.
\begin{figure}[!h]
\begin{center}
 \includegraphics[scale=0.40]{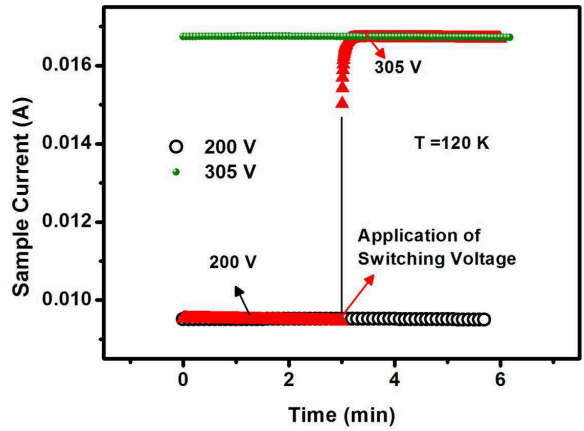}
 \end{center}
\caption{(color on line) Time dependent current measurement on application of switching voltage for LaMnO$ _{3} $-2.}
\end{figure}
We apply 200 V in the sample for nearly 3 minutes and then suddenly step the bias to 305 V and we observe that the sample current switches to higher level but stays constant with the time.\\
We have also done a calculation for the maximum temeprature rise in the sample due to the Joule heating considering the heat generated in the sample and dissipated from the sample through the metal base via the GE-Varnish and thermal grease. \\
The heat balance equation in the sample is given as:
\begin{equation}
 C_{P}\frac{dT}{dt} = \frac{d}{dt}(Q_{in}+ Q_{out})
 \end{equation}
 \begin{equation}
 C_{P}\frac{dT}{dt} = IV-C_{P}(\frac{T-T_{S}}{\tau})
 \end{equation}
 Where T$ _{S} $ is the base temperature,\\
 $ \tau $ = Thermal relaxation time betwen the samle and the base,\\
 $ \tau $ = C$ _{P} $R$ _{Th} $; C$ _{P} $ is the heat capacity and R$ _{Th} $ is the thermal boundary resistance,\\
  and P= IV.
 \begin{equation}
 \frac{dT}{dt} = \frac{P}{C_{P}}-\frac{T}{\tau}+\frac{T_{S}}{\tau}; 
 \end{equation}
 \begin{equation}
 \frac{dT}{dt}+ bT-(\frac{P}{C_{P}}+bT_{S});
 \end{equation}
 
 Soution to this equation is given as
 \begin{equation}
 T(t) = T_{S}+ \frac{P\times\tau}{C_{P}}(1- \exp (-\frac{t}{\tau}))
 \end{equation}
 
Using measured value of C$ _{P} $ and R$ _{Th} $, we evaluate the evolution of T(t) on application of the pulsed power \cite{Himanshu,Lakeshore}. A typical result of temperature rise $ \Delta $ T= T(t)-T$ _{S} $ as function of time is shown in Figure 10 at T$ _{S} $=150K for LaMnO$ _{3} $-1.\\
\begin{figure}[!h]
   \begin{center}
  \subfigure[] {\includegraphics[scale=0.33]{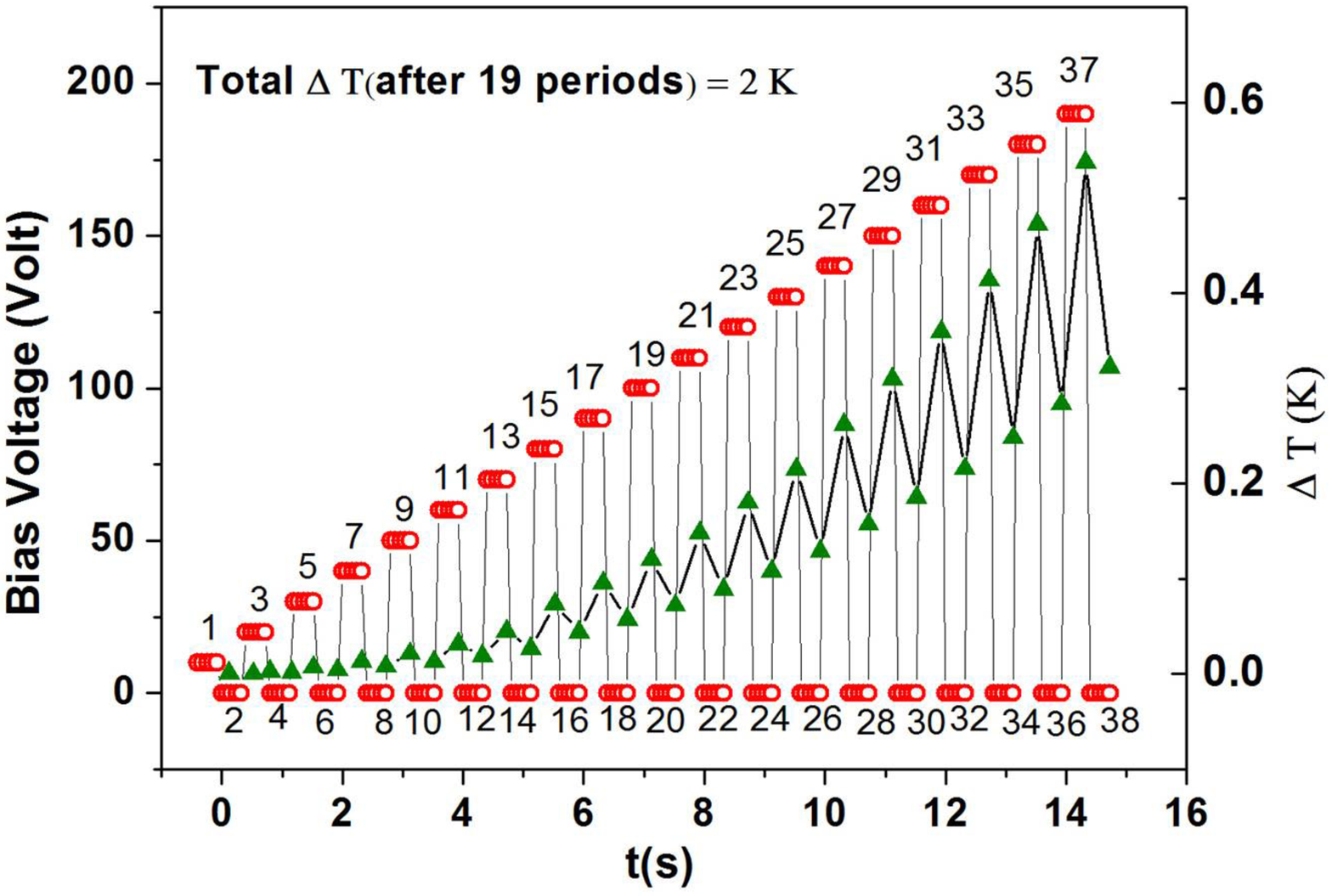}}
   \subfigure[] {\includegraphics[scale=0.33]{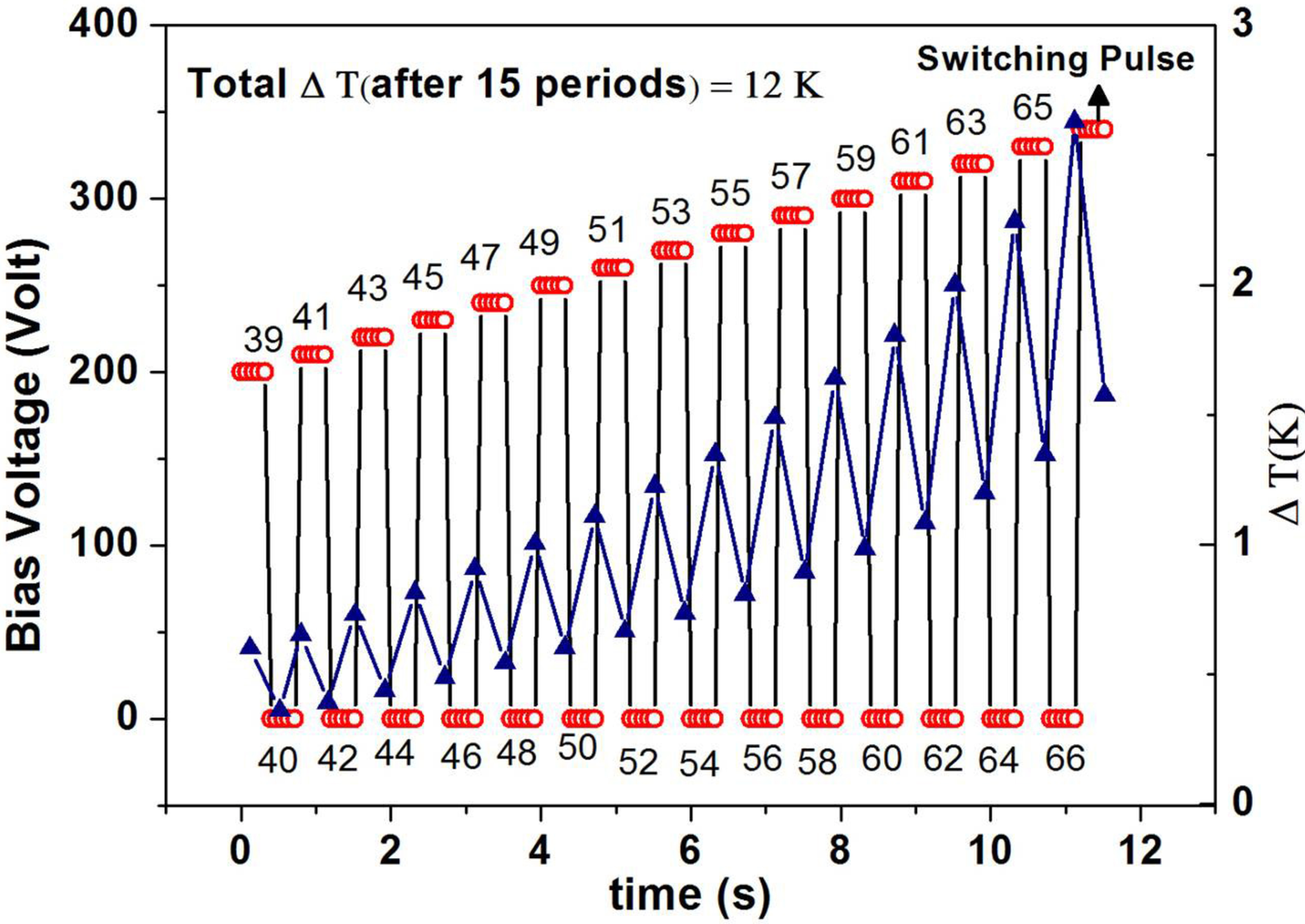}} 
   \end{center}
   \caption{(color on line) (a)Temperature rise and fall ($ \Delta $ T) for each ON and OFF pulses from 10 to 190 V and (b) from 200 to 340 V.  Total increment in temperature is calculated by adding the $ \Delta $ T after each period i.e. for pulse no. 2,4, 6.......68 indicated in the Fig at each pulse position.}
   \end{figure}
 The highest temerature rise at the maximum power pulse of nearly 300 mW is $ < $ 3K (switching pulse in Figure 10(b).
 In a similar way calculating the $ \Delta $ T for each pulses from 10 V to 340 V, we find that the total  temperaure rise is $ \sim $ 14K (from Figure 10(a) and 10(b)). This tempearture rise (maximum estimated) is not sufficent to generate such kind of resistive transition which we have observed at 150K.
 We find that the temperature rise even at the lowest temperature is $ \leq $ 4K with a maximum power applied. This rules out any substantial contribution to the result arising from Joule heating. In general, when the applied voltage across the sample is increased in a step, so that switching in resistive state occurs, the resistance in response to the step may drift due to Joule heating. We have carried out such test and find absence of such drift. We have also showed that when data are taken with a constant bias the material makes a sharp transition from a lower resistance state to a higher resistance state at low temperature (in Figure 5) which is opposite to what one expects if the entire effect is due to Joule heating. 
\section{\bf SUMMARY}
In summary, we observe a sharp transition in the resistive state of nominally pure LaMnO$_{3}$ at temperatures below 300K with a moderate applied bias. The field induced resistive state transition seen in this report is different from those seen in manganites with higher level of Mn$^{4+}$ content. Thus the mechanisms proposed for them are not applicable for the observations made here. In this case the transition occurs between the polaronic insulating state in LaMnO$_{3}$ and a bad metallic phase that has a temperature independent resistivity. The experiments were carried out on three single crystals. All the three crystals, despite differences in absolute values of resitivities, show qualitatively similar bias induced resistive transitions.  At lower temperatures change at the transition can be as high as four orders of magnitude. The transition becomes softer and eventually not observed above 280K.  Similar experiments were also carried out on single crystal of Sr substituted La$ _{0.9} $Sr$ _{0.1} $MnO$ _{3}$. In this case we find that the increase of Mn$^{4+}$ content on Sr substitution inhibits the resistive state transition. 

The observation has been explained as bias driven percolation type transition between the two coexisting phases mentioned above.  The applied bias can change the mobile fraction $f$ of the bad metallic phase leading to a percolation type transition to the lower resistive state when the fraction $f$ crosses a critical volume fraction for percolation transition. The mobile fraction $f$ (as estimated from the experimental data) has an activated dependence on temperature with activation energy $\approx$ 200meV. The fraction $f$ has a dependency on the field $f =C exp \frac{-E_{0}}{E}$ . This leads to rapid enhancement of $f$ on application of field lading to a sharp transition in the resistive states. 

Likely microscopic mechanisms for co-existing phases have been proposed based on recent suggestions for appearance and existence such phases. It appears that a bad metallic minority phase can co-exist in nominally pure LaMnO$_3$ due to charge transfer instability driven charge disproportionation\cite{Moskvin}. 

\section{Acknowledgments}
AKR and RN acknowledge the financial support from the Department of Science and Technology, Government of India as a sponsored project. RN acknowledges support from CSIR as Senior Research Fellowship. YM nad AKR acknowledge support from the joint DST-RFBR project (N 11-02-92706, DNT grant).


\section*{References}
\begin{thebibliography} {100}
\bibitem{CNR} C. N. R. Rao and B. Raveau 1998 {\it Colossal Magnetoresistance, Charge Ordering and Related Properties of Manganese Oxides, World Scientific}.
\bibitem{Tokura}Y. Tokura 1999 {\it Colossal-Magnetoresistive Oxides,Gordon and Breach Science Publishers, New York}.
\bibitem{Goodenough}John B. Goodenough and J.-S. Zhou 2007 {\it J. Mater. Chem.} {\bf 17} 2394.
\bibitem{Rodriguez}J. Rodriguez-Carvajal, M. Hennion, F. Moussa, A. H. Moudden, L. Pinsard and A. Revcolevschi 1998 {\it Phys. Rev. B} {\bf 57} 3189.
\bibitem{BBvanAken}B. B. van Aken, A. Meetsma, Y. Tomioka, T. Tokura, and T. T. M. Palstra 2003
{\it Phys. Rev. Lett.}{\bf 90} 66403.
\bibitem{TChaterji}T. Chaterji, B. Ouladdiaf, P. Mandal, B. Bandyopadhyay, and B. Ghosh 2002
{\it Phys.  Rev.  B} {\bf 66} 054403.
\bibitem{Loa} I. Loa, P. Adler, A. Grzechnik, K. Syassen, U. Schwarz, M. Hanfland, G. Kh. Rozenberg,
P. Gorodetsky, and M. P. Pasternak 2001 {\it Phys. Rev. Lett.} {\bf 87} 125501.
\bibitem{Brion} S. de Brion, F. Ciorcas, and G. Chouteau, P. Lejay, P. Radaelli, C. Chaillout 1999
{\it Phys.  Rev. B} {\bf 59} 1304.
\bibitem{Dipten} Parthasarathi Mondal and Dipten Bhattacharya, P. Mandal 2011 {\it Phys. Rev. B} {\bf 84} 075111.
\bibitem{CNR2} C. N. R. Rao, A. R. Raju, V. Ponnambalam, and Sachin Parashar, N. Kumar 2000 {\it Phys. Rev. B} {\bf 61} 594.
\bibitem{Tokura2} A. Asamitsu, Y. Tomioka, H. Kuwahara and Y. Tokura 1997 {\it Nature} {\bf 388} 50.
 \bibitem{AyanG1} Ayan Guha, Arindam Ghosh and A. K. Raychaudhuri, S. Parashar, A. R. Raju, and C. N. R. Rao 1999 {\it App. Phys. Lett.} {\bf 75} 3381.
 \bibitem{AyanG} Ayan Guha and A. K. Raychaudhuri, A. R. Raju and C. N. R. Rao 2000 {\it Phys. Rev. B} {\bf 62} 5320.
\bibitem{Himangshu} Himanshu Jain, A. K. Raychaudhuri, Ya. M. Mukovskii and D. Shulyatev 2006 {\it App. Phys. Lett.} {\bf 89} 152116.
\bibitem{Aveek}  Aveek Bid, Ayan Guha and A. K. Raychaudhuri 2003 {\it Phys. Rev. B} {\bf 67} 174415.
\bibitem{Rickard Fors} Rickard Fors, Sergey I. Khartsev, and  Alexander M. Grishin 2005 {\it Phys.  Rev.  B}  {\bf 71} 045305.
 \bibitem{Shohini} Sohini Kar and A. K. Raychaudhuri 2007 {\it App.  Phys.  Lett.} {\bf 91} 143124.
\bibitem{Jon-Olaf} Jon-Olaf Krisponeit, Christin Kalkert, Bernd Damaschke, Vasily Moshnyaga, and Konrad Samwer 2010 {\it Phys. Rev. B} {\bf 82} 144440.
\bibitem{sawa} A. Sawa, T. Fujii, M. Kawasaki, and Y. Tokura 2004 {\it App.  Phys.  Lett.} {\bf 85} 4073.
\bibitem{Nian} Y. B. Nian, J. Strozier, N. J. Wu, X. Chen, and A. Ignatiev 2007 {\it Phys.  Rev.  Lett.} {\bf 98} 146403.
 \bibitem{Himangshu2} Himanshu Jain and A. K. Raychaudhuri 2008 {\it App.  Phys. Lett.} {\bf 93} 182110.
 \bibitem{Tokunaga} M. Tokunaga, Y. Tokunaga, and T. Tamegai 2004 {\it Phys.  Rev.  Lett.} {\bf 93} 037203.
  \bibitem{Jardim} A. S. Carneiro, R. F. Jardim, and F. C. Fonseca 2006 {\it Phys.  Rev.  B} {\bf 73} 012410. 
 \bibitem{ChJooss} Ch. Jooss, L.Wu, T. Beetz, R. F. Klie, M. Beleggia, M. A. Schofield, S. Schramm, J. Hoffmann, and Y. Zhu 2007 {\it PNAS} {\bf{104}} 13601.
 \bibitem{Mukovski2} D. Shulyatev, S. Karabashev, A. Arsenov, Ya. Mukovskii, S. Zverkov 2002, {\it J.  Cryst.  Growth} {\bf 237-239} 810.
  \bibitem{Himangshu3} Himanshu Jain, A.K. Raychaudhuri, Ya.M. Mukovskii, D. Shulyatev 2006 {\it Solid State Communications} {\bf 138} 318.
 \bibitem{Mukovski} A. de Andrés, N. Biškup, M. García-Hernández,and Y. M. Mukovskii 2009 {\it Phys.  Rev.  B} {\bf 79} 014437.
  \bibitem{Renard}A-M Haghiri-Gosnet and J-P Renard 2003 {\it J. Phys. D: Appl. Phys.} {\bf 36} R127-R150. 
  \bibitem{Dagotto} Elbio DAGOTTO, Takashi HOTTA, Adriana MOREO 2001 {\it Physics Reports} {\bf 344} 1-153.  
  \bibitem{Qingzhong} Qingzhong Xue 2003 {\it Physica B} {\bf 325} 195.
 \bibitem{Bardeen} John Bardeen 1980 {\it Phys.  Rev.  Lett.} {\bf 45} 1978.
  \bibitem{Farges} J. P. Farges  1994 {\it Organic Conductors, Marcel Dekker, Inc. New York}.
 \bibitem{Huang} Q. Huang, A. Santoro, J. W. Lynn, R. W. Erwin, J. A. Borchers, J. L. Peng, and R. L. Greene 1997 {\it Phys.  Rev.  B} {\bf 55} 14987.
  \bibitem{Prado} F. Prado, R. Zysler, L. Morales, A. Caneiro, M. Tovar, and M. T. Causa 1999 {\it J.  Magn. Magn.  Mater.} {\bf 196} 481.
 \bibitem{Kovaleva} N. N. Kovaleva, A. V. Boris, C. Bernhard, A. Kulakov, A. Pimenov, A. M. Balbashov, G. Khaliullin and B. Keimer 2004 {\it Phys.  Rev.  Lett.} {\bf 93} 147204.
 \bibitem{Moskvin} A. S. Moskvin 2009 {\it Phys.  Rev.  B} {\bf 79} 115102.
 \bibitem{Ramos} A. Y. Ramos,  N. M. Souza-Neto,  H. C. N. Tolentino,  O. Bunau,  Y. Joly,  S. Grenier, J.-P. Itié,  A.-M. Flank,  P. Lagarde  and  A.  Caneiro 2011 {\it EPL} {\bf 96} 36002.
 \bibitem{Murakami} Y. Murakami,  J. P. Hill,  D. Gibbs,  M. Blume,  I. Koyama,  M. Tanaka,  H. Kawata,  T. Arima, Y. Tokura, K. Hirota and Y. Endoh 1998 {\it Phys.  Rev.  Lett.} {\bf 81} 582.
 \bibitem{Zimmermann} M. V. Zimmermann, C. S. Nelson, Y. J. Kim, J. P. Hill, D. Gibbs, H. Nakao, Y. Wakabayashi, Y. Murakami, Y. Tokura, Y. Tomioka, T. Arima, C.-C. Kao, D. Casa, C. Venkataraman, and Th. Gog 2001 {\it Phys. Rev. B} {\bf 64} 064411.
\bibitem{Himanshu}Himanshu Jain 1998 {\it An investigation of the ferromagnetic insulating state
of manganites, PhD thesis, Indian Institute of Science, Bangalore}.
\bibitem{Lakeshore} Lakeshore accesories, www.lakeshore.com.
\end{thebibliography}
\end{document}

\title{Supplementary}
\author{Rajib Nath$^1$, A. K. Raychaudhuri$^1$, Ya. M. Mukovskii$^2$, Parthasarathi Mondal$^3$, Dipten Bhattacharya$^3$ and P. Mandal$^4$}
\address{$^1$ Department of Condensed matter physics and Materials Science, S.N. Bose National Center for Basic Sciences, Kolkata 700098, India}
 \address{$^2$ National Research Technological University,(MISIS),Leninskii prosp. 4, Moscow, 119049, Russia}
 \address{$^3 $ Nanostructured Materials Division, CSIR-Central Glass and Ceramic Research Institute, Kolkata 700032, India}
  \address{$^4$ Experimental Condensed Matter Physics, Saha Institute of Nuclear Physics, Kolkata 700064, India}
\ead{rajibnath.bu@gmail.com} 
\ead{arup@bose.res.in }
\submitto{\JPCM}
\maketitle
The experimental measurements have been carried out on three LaMnO$ _{3} $ single crystals and one La$ _{0.9} $Sr$_{0.1}$MnO$ _{3} $ single crystals and the dimensions are given below. 
LaMnO$ _{3} $-1 $\rightarrow$ Diameter: 5 mm and Thickness: 1 mm. Distance between +I $ \leftrightarrow $ -I pad = 3 mm and 
+V$ \leftrightarrow $ -V pad = 0.6 mm.
LaMnO$ _{3} $-2 $ \rightarrow $ Diameter: 4 mm and Thickness: 1mm. Distance between +I $ \leftrightarrow $ -I pad = 0.75 mm and 
+V $ \leftrightarrow $ -V pad = 0.2 mm.\\
La$ _{0.97} $MnO$_{3}$ marked as LaMnO$ _{3} $-3 $ \rightarrow $ Diameter: 3 mm and thickness: 1 mm. Distance between +I $ \leftrightarrow $ -I pad = 2.5 mm and +V $ \leftrightarrow $ -V pad = 0.3 mm.\\  La$_{0.9}$Sr$_{0.1}$MnO$ _{3}$ $ \rightarrow $ Diameter: 4 mm and thickness: 1 mm. Distance between +I $ \leftrightarrow $ -I pad = 0.8 mm and  +V $ \leftrightarrow $ -V pad = 0.4 mm.
\section{\bf Charecterisation of the sample}
\subsection {High temperature XRD}
\begin{figure}[h]
 \begin{center}
   \subfigure[]{\includegraphics[scale=0.34]{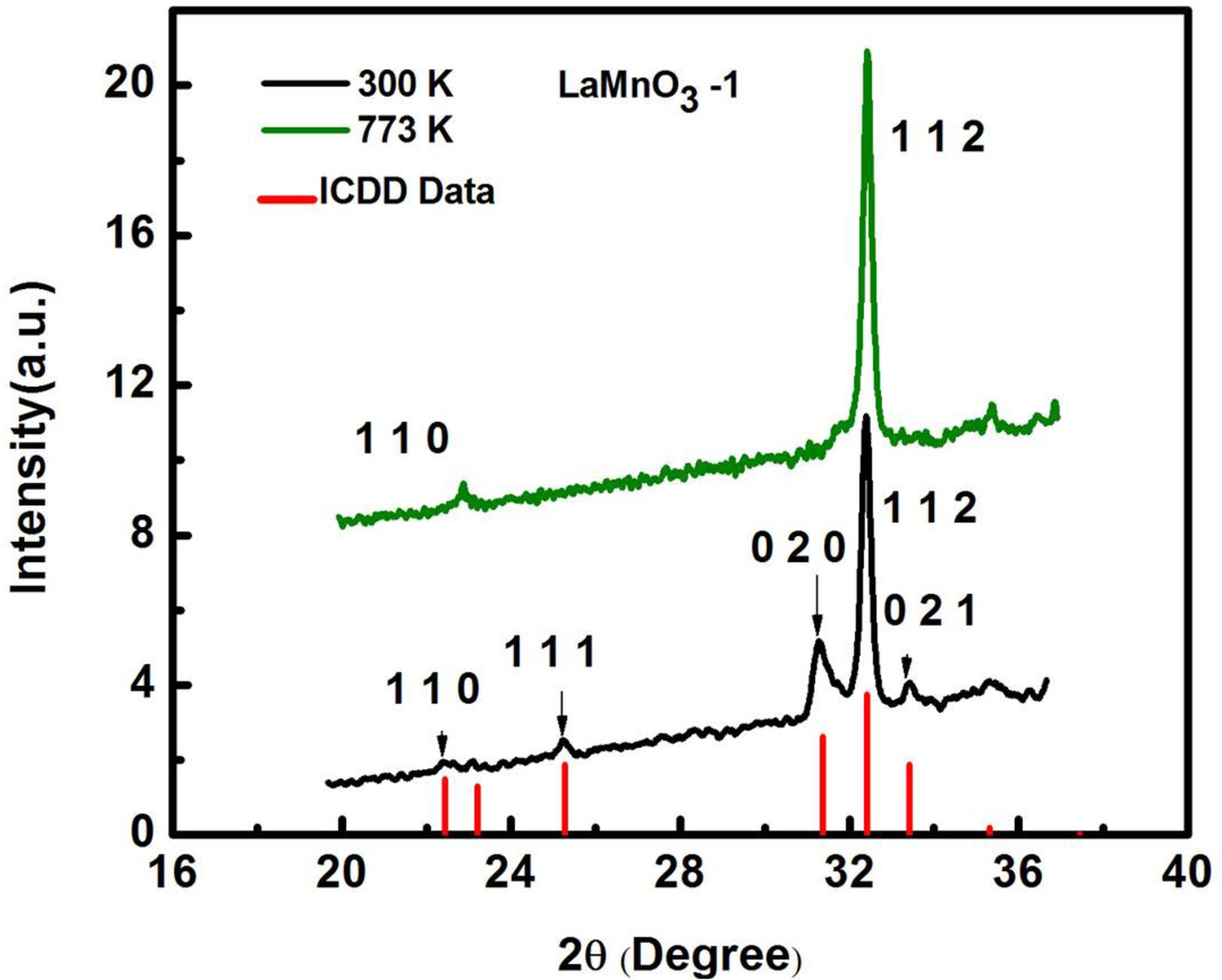}}
   \subfigure[]{\includegraphics[scale=0.28]{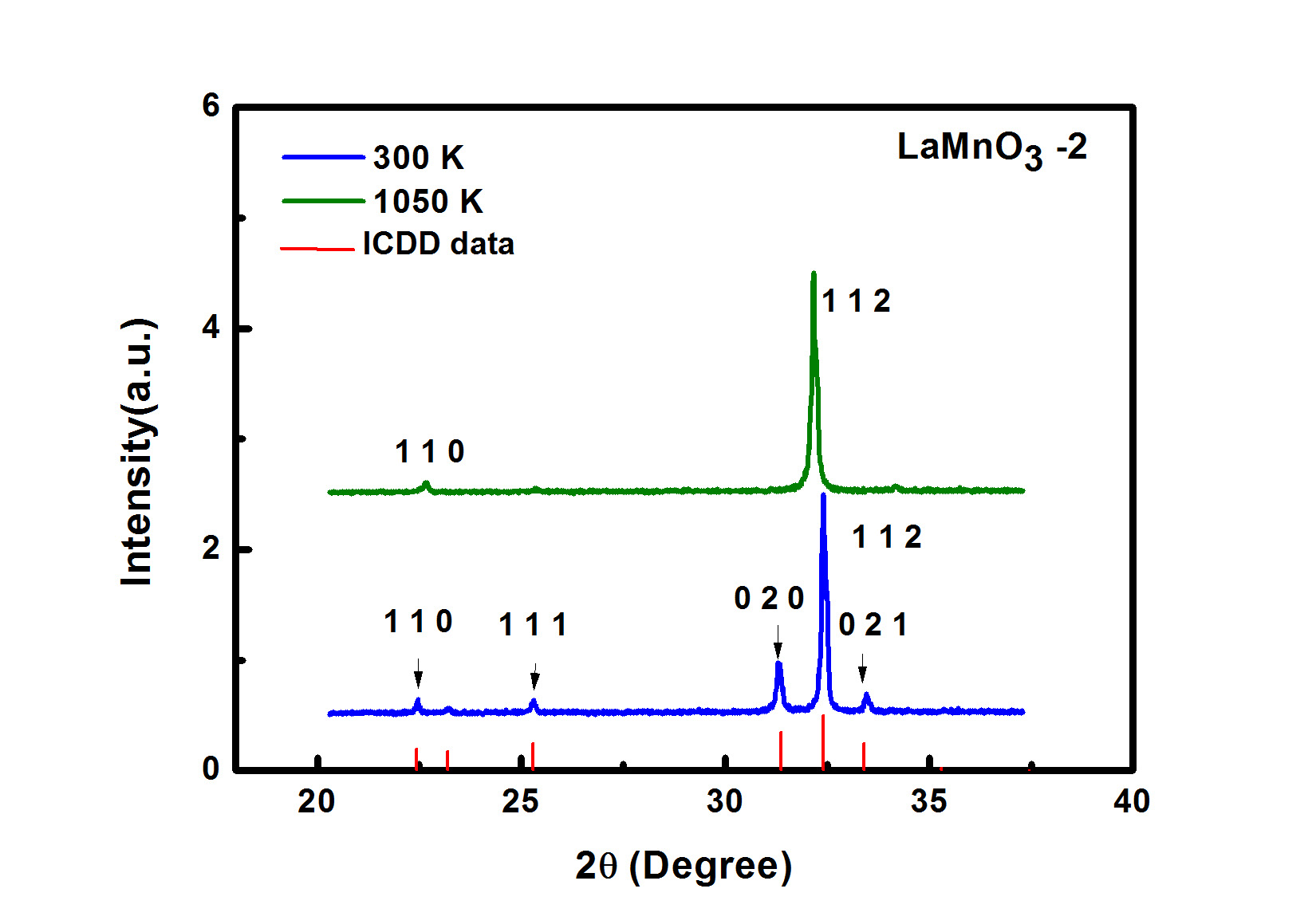}}
    \subfigure[]{\includegraphics[scale=0.28]{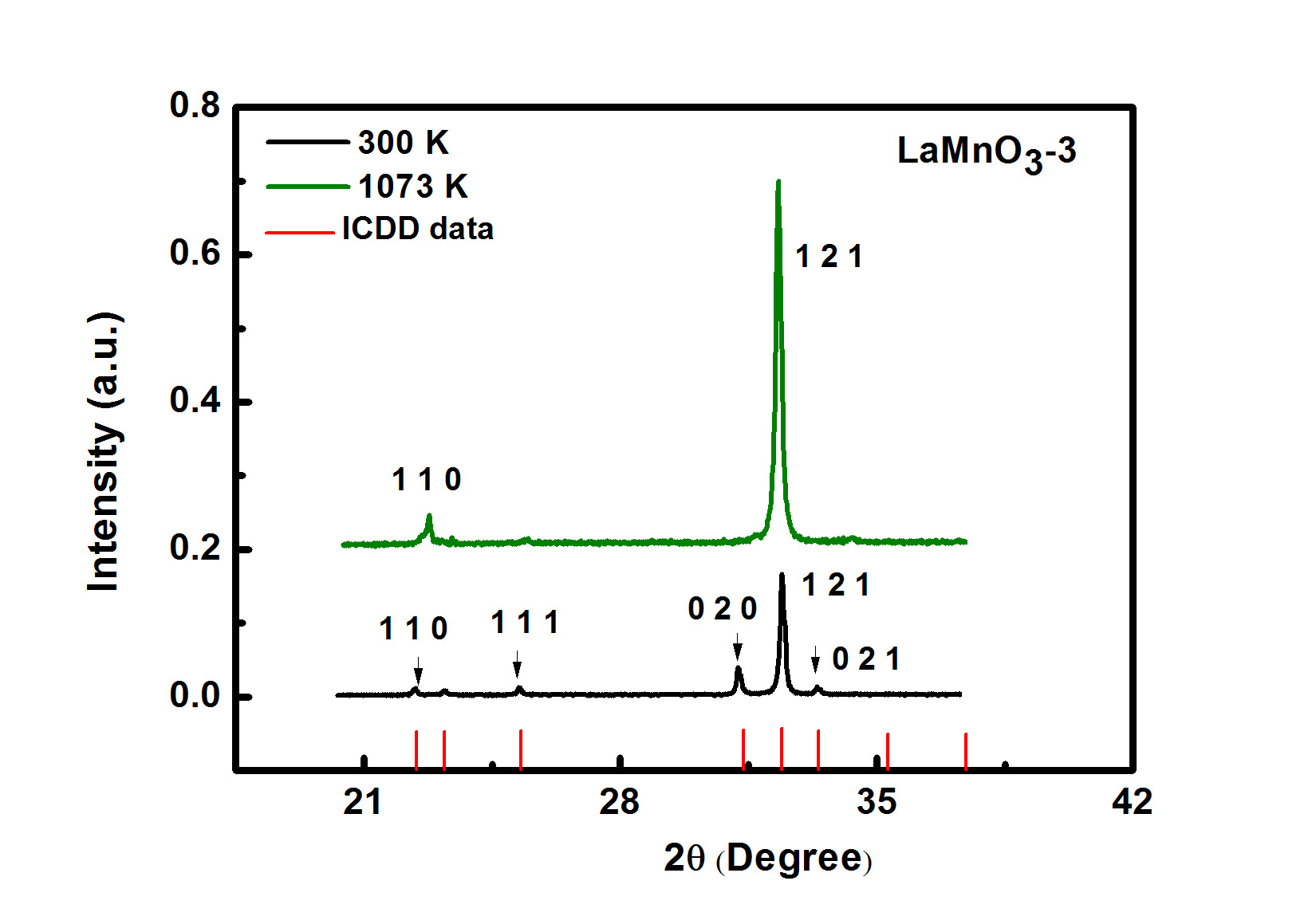}}
 \end{center}
   \caption{(color on line) (a),(b) and (c) High temperature XRD of three LaMnO$_{3}$ single crystals.}
 \end{figure} 
Nominally pure LaMnO$ _{3} $ has a John-Teller distorted orthorhombic structure at 300K. The John-Teller distortion and orthorhombic structure disappeared at nearly 750K. This transition temperature is reduced with the presence of Mn $ ^{4+} $ in the sample.
High tempearture XRD is an useful method to check the structural transition and also stoichiometry of the parent LaMnO$_{3} $. We have done the high temperature XRD of our three LaMnO$ _{3}$ samples (crushing the single crystal into powder) from  $ 25^{0}C $ to $ 950^{0}C $ and we find that all the samples shows their JT distortion tempearture above 750K which indicates nearly perfect LaMnO$ _{3} $ structure of our samples (in Figure 1(a),1(b) and 1(c)). However, small variation in the transition temperature reflects the slight difference in Mn$ ^{4+} $ content and as well as small differences in oxygen content in the three crystals.
\section{\bf Electrical measurement of the samples}
We have done four probe I-V measurements on all the three LaMnO$_{3}$ samples and observe the same effect of electric field driven resistive transition in other two samples (LaMnO$ _{3} $-2 and LaMnO$ _{3} $-3) as LaMnO$ _{3} $-1 (in Figure 2(a) and 2(b)).
\begin{figure}[h]
  \begin{center}
   \subfigure[]{\includegraphics[scale=0.55]{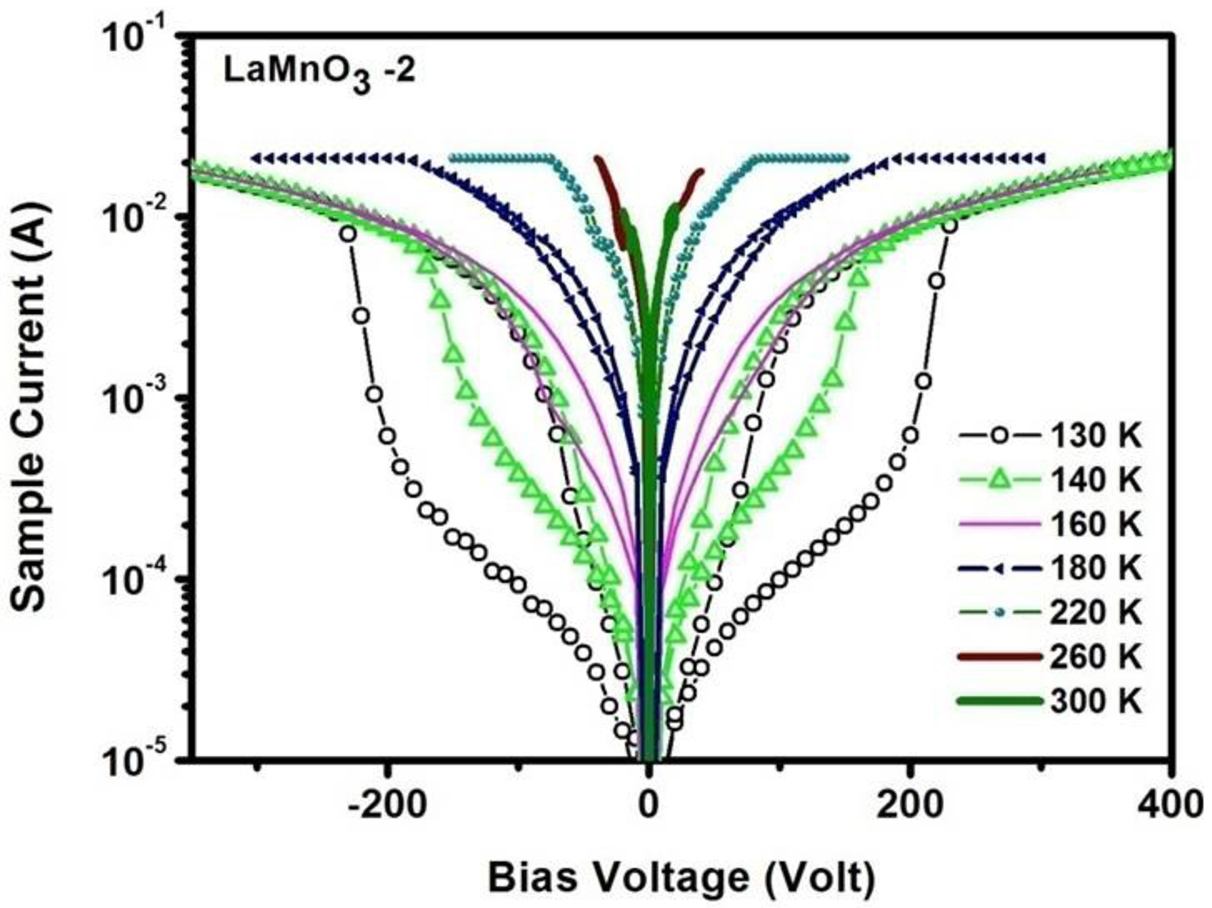}} 
   \subfigure[]{\includegraphics[scale=0.55]{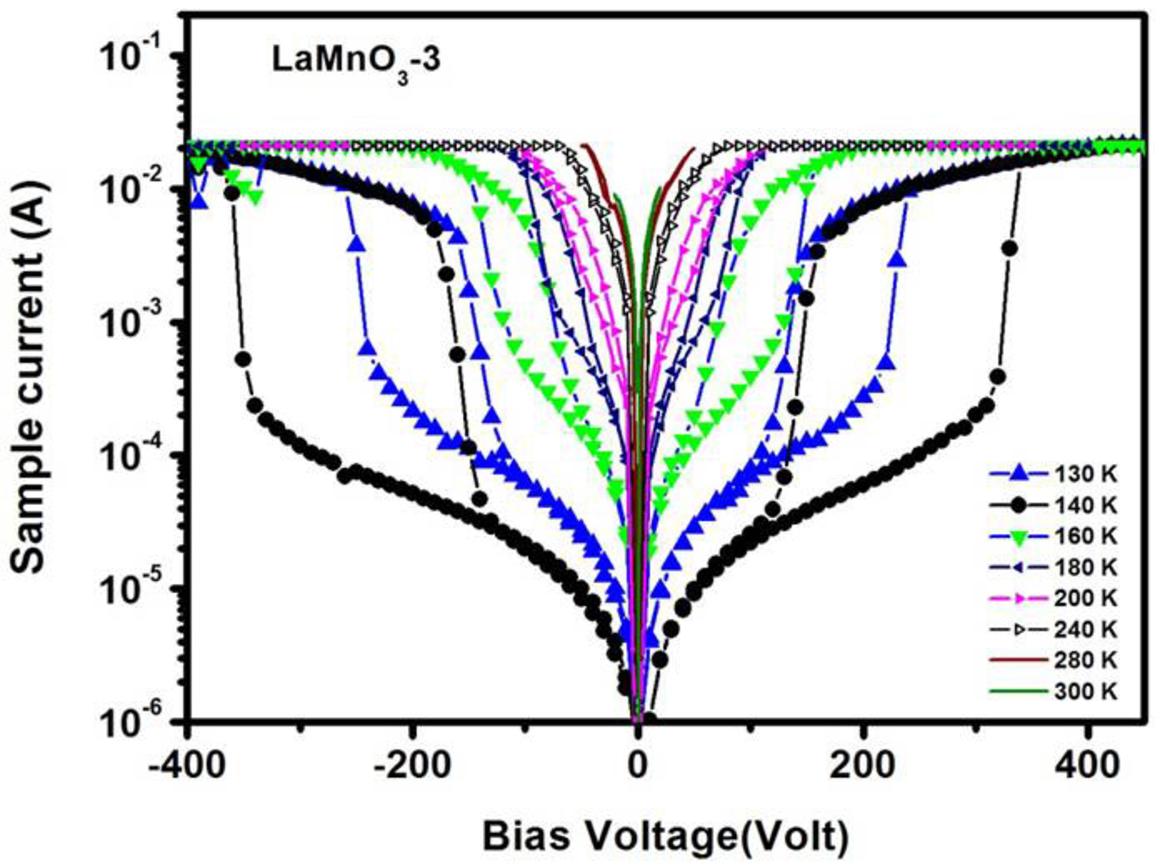}}\\
   \subfigure[]{\includegraphics[scale=0.38]{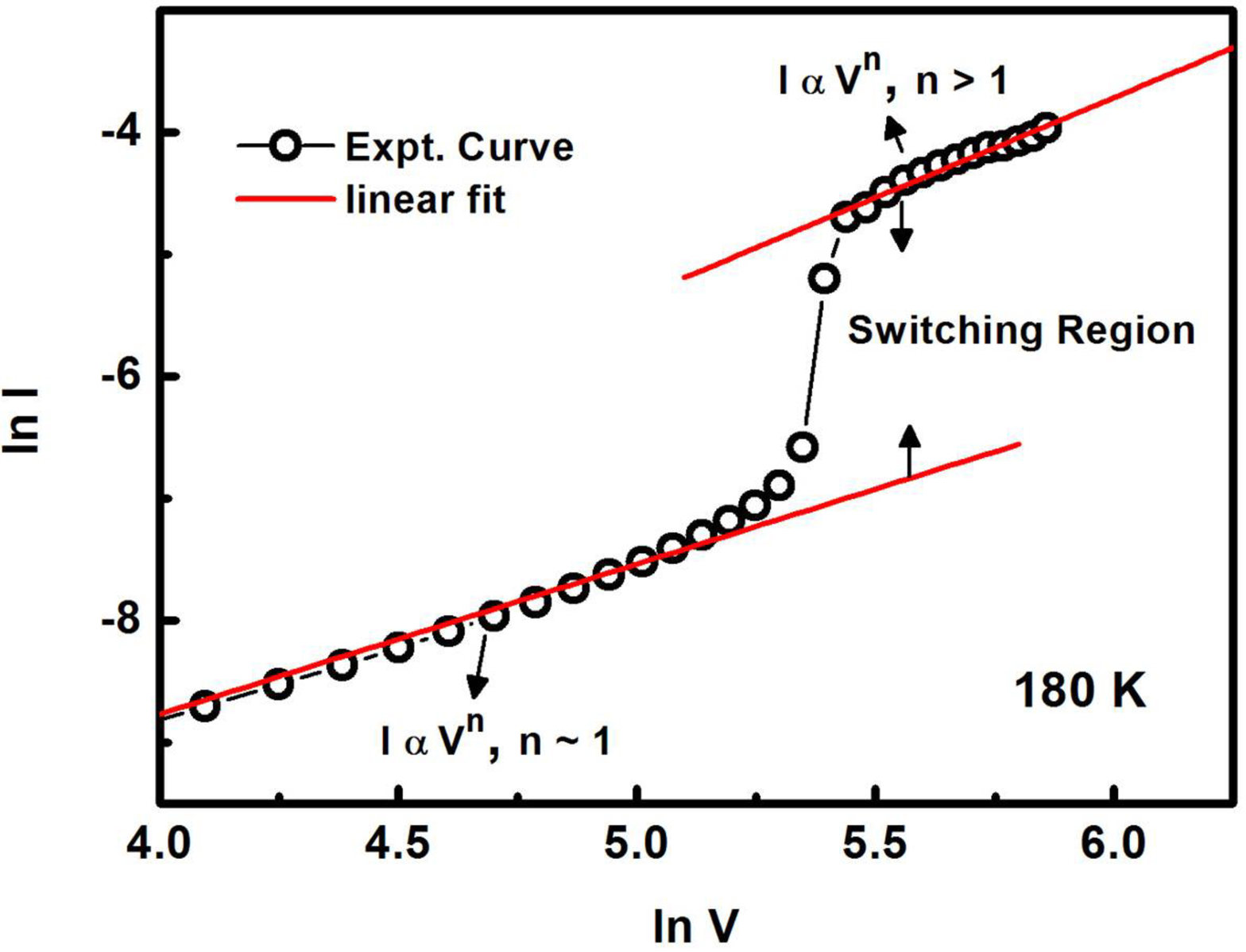}}
   \end{center}
   \caption{(color on line) (a),(b) $ I-V $ curves for the LaMnO$ _{3} $-2 and LaMnO$ _{3} $-3 and (c) Power law dependence of $ I-V $ curve before and after switching at 180K.}
  \end{figure}
We find a power law ($ I\propto V^{n} $) dependence of the  $ I-V $ curves before and after the transition at 180K shown in Figure 2(c).\\
\begin{figure}[h]
\begin{center}
\subfigure[]{\includegraphics[scale=0.55]{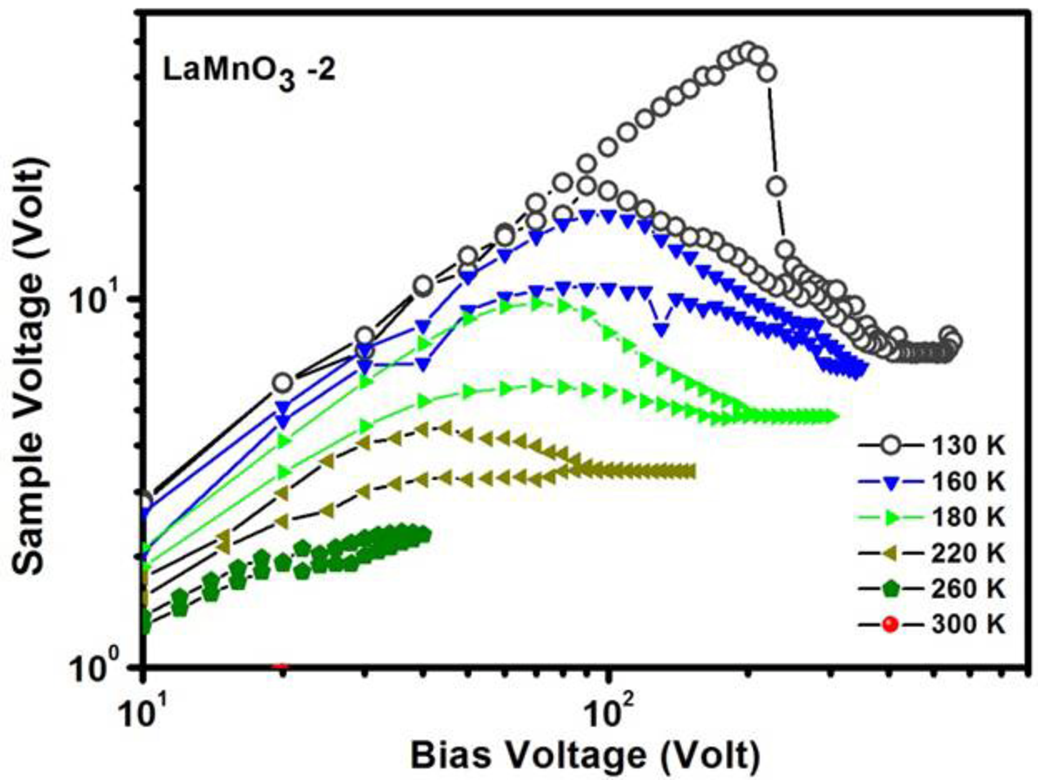}}
\subfigure[]{\includegraphics[scale=0.50]{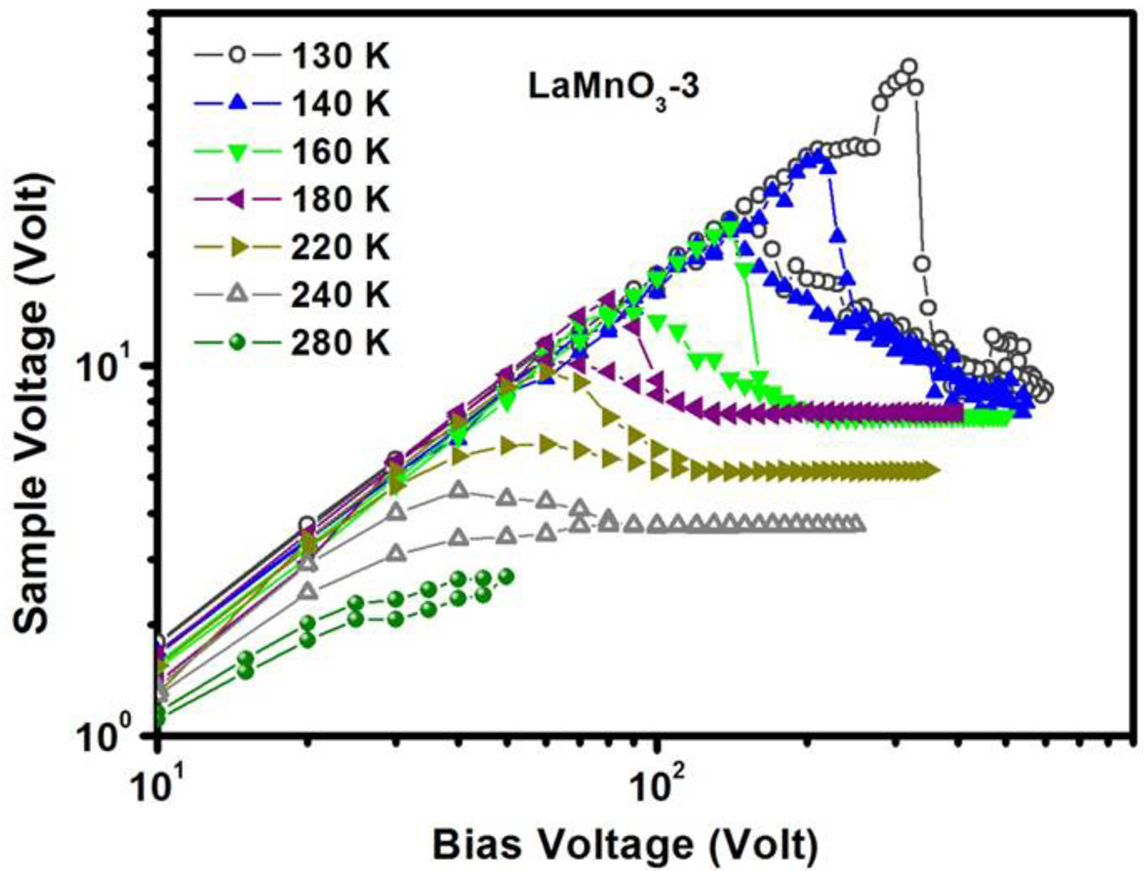}}
\end{center}
 \caption{(color on line) (a),(b)Sample voltage (V) vs. Bias Voltage (V$ _{appl} $) curves for LaMnO$ _{3} $-2 and LaMnO$ _{3} $-3.}
\end{figure}
Sample voltage behaviour with the applied bias in LaMnO$ _{3} $-2 and LaMnO$ _{3} $-3 is shown in Figure 3(a) and 3(b).